\documentclass[pdflatex,sn-mathphys-num]{sn-jnl}

\usepackage{graphicx}%
\usepackage{multirow}%
\usepackage{amsmath,amssymb,amsfonts}%
\usepackage{amsthm}%
\usepackage{mathrsfs}%
\usepackage[title]{appendix}%
\usepackage{xcolor}%
\usepackage{textcomp}%
\usepackage{manyfoot}%
\usepackage{booktabs}%
\usepackage{algorithm}%
\usepackage{algorithmicx}%
\usepackage{algpseudocode}%
\usepackage{listings}%
\usepackage{bm}%

\theoremstyle{thmstyleone}%
\theoremstyle{thmstyletwo}%

\theoremstyle{thmstylethree}%

\begin{document}

\title{Theory of hybrid defects, with coupled orientational order parameters, on flat and curved surfaces}


\author[1]{\fnm{Lincoln} \sur{Paik}}\email{lpaik@kent.edu}

\author*[1]{\fnm{Jonathan V.} \sur{Selinger}}\email{jselinge@kent.edu}

\affil*[1]{\orgdiv{Department of Physics, Advanced Materials and Liquid Crystal Institute}, \orgname{Kent State University}, \orgaddress{\city{Kent}, \state{OH} \postcode{44242}, \country{USA}}}


\abstract{Many physical systems involve two types of orientational order, which are coupled together. For example, ferroelectric nematic liquid crystals have coupled polar and nematic order, and tilted hexatic phases have coupled polar and hexatic order. In these systems, defect structures can be quite complex. Here, we investigate phases with two types of two-dimensional orientational order, $m$-atic and $n$-atic, where $m$ and $n$ are two distinct integers. We simulate these phases in a flat disk with strong radial anchoring, and on a spherical surface, because both of these geometries require the presence of defects. If the coupling between the two types of order is weak, then the defects are connected by a network of diffuse walls, and the system forms a stable domain structure. As the coupling increases, the domain walls become sharper and shorter. For very strong coupling, the higher-order defects merge into the lower-order defects, forming stretched defect cores.}




\maketitle

\section{Introduction}

Topological defects are fundamental features of ordered phases in condensed matter physics, as discussed in several review articles and textbooks~\cite{Mermin1979,Kleman1983,Chaikin1995,Kleman2003,Alexander2012,Selinger2024}.  These defects play essential roles in the structure, dynamics, and statistical mechanics of hard and soft matter.  Furthermore, observation of defects is often a useful method for recognizing ordered phases.

Many physical systems have different types of orientational order.  This order may be \emph{polar}, with one-fold rotational symmetry, as in ferromagnets.  It may be \emph{nematic}, with two-fold rotational symmetry, as in the most common phases of liquid crystals.  It may also have a higher-order rotational symmetry, such as a \emph{tetratic} phase with four-fold symmetry, or a \emph{hexatic} phase with six-fold symmetry~\cite{Nelson1979}.  In general, researchers use the term $n$-atic to refer to orientational order with $n$-fold rotational symmetry.  In any phase with $n$-atic order, topological defects are points in two dimensions (2D).  Around a point defect, the orientational order rotates through an angle of $2\pi/n$ radians.

In certain cases, a physical system might have more than one type of orientational order.  For example, a liquid crystal might have strong nematic order (all molecules aligned up or down along a certain axis), along with weak polar order (51\% of molecules pointing up and 49\% of molecules pointing down along that axis).  Alternatively, a thin film of liquid crystal might have hexatic order (six-fold symmetry in the orientations of intermolecular bonds), along with polar order (one-fold symmetry in the directions of molecular tilt)~\cite{Nelson1980,Selinger1988,Selinger1989}.  In general, there might be a combination of $m$-atic and $n$-atic order, where $m$ and $n$ are two distinct integers.

In a phase with two types of orientational order, what happens to the topological defects?  This question was addressed in two important papers in the 1980's.  In a theoretical paper, Lee and Grinstein considered the general concept of a phase with 2D polar and nematic order~\cite{Lee1985}.  They showed that this system forms ``string defects,'' each consisting of two $\pm1/2$ point defects in the nematic order, connected by a domain wall in the polar order.  Around the same time, Dierker, Pindak, and Meyer investigated liquid-crystal films with 2D polar (tilt) and hexatic order~\cite{Dierker1986}.  They found experimentally and explained theoretically that these films form ``star defects,'' each with five $1/6$ point defects in the hexatic order at the tips of a star and one more at the center, connected by tilt domain walls.  A further theoretical study considered how the star defects should evolve through transitions between phases with different couplings between tilt and hexatic order~\cite{Selinger1989}.  The general concept of string defects and star defects is reviewed in Section 7.2 of the textbook~\cite{Selinger2024}.

In this paper, we further explore the structure of topological defects in phases with different types of orientational order that are coupled together.  We will refer to this general class of defects as ``hybrid defects''~\cite{Yi2025}.  There are several motivations for this study:

First, the ferroelectric nematic phase has recently been discovered in newly synthesized liquid crystals~\cite{Nishikawa2017,Chen2020,Lavrentovich2020,Sebastian2022,Mandle2022}. Unlike the conventional nematic phase, the ferroelectric nematic phase has polar order of the molecules, leading to an electrostatic polarization.  Hence, these materials have a combination of polar and nematic order, similar to the combination studied theoretically by Lee and Grinstein~\cite{Lee1985}, although the electrostatic interaction is more complex than the energy function in their model. Experiments have observed topological defects in ferroelectric nematic liquid crystals, and they have interesting extended structures, involving both polar and nematic order~\cite{Basnet2022,Kumari2023,Yi2025}.

Second, simulations by Glotzer and collaborators have studied how particles pack to form a colloidal crystal on the surface of a sphere~\cite{Jones2025}.  The spherical geometry is important because it requires the presence of topological defects with a total topological charge of +2.  In these simulations, the defect structure depends on the shape of the particles.  If the particles are rounded cubes, they pack in a lattice with four-fold symmetry.  The tetratic orientational order leads to eight defects, each with topological charge +1/4, around the sphere.  By contrast, if the particles are rounded tetrahedra, they form a more complex woven motif, which we might interpret as a combination of nematic and tetratic orientational order.  In that case, the simulations show a network of extended defects and domain walls around the sphere.

Third, models of active matter often exhibit a combination of polar and nematic order.  In some cases, this combination leads to string defects, which have two +1/2 defects in the nematic order connected by a domain wall in the polar order~\cite{Mishra2025,Dinelli2026}.  The presence of these defects controls the dynamic evolution of the active material.

Fourth, researchers have recently analyzed the arrangement of biological cells in epithelial tissue~\cite{Armengol-Collado2023,Happel2025}.  These analyses show that the arrangement has a combination of different types of $n$-atic orientational order, particularly of nematic and hexatic order.  This combination has at least the potential to generate hybrid defects.

Fifth, Vafa and Doostmohammadi have recently published a theory for string defects in ``nematopolar'' phases with combined polar and nematic order~\cite{Vafa2025}.  They argue that these defects lead to a new universality class for the transition between nematic and nematopolar phases.

Here, we consider the general phenomenon of hybrid defects involving two distinct types of orientational order, \hbox{$m$-atic} and \hbox{$n$-atic}.  We explore four combinations that commonly occur:  (a)~polar ($m=1$) and nematic ($n=2$), (b)~nematic ($m=2$) and tetratic ($n=4$), (c)~polar ($m=1$) and hexatic ($n=6$), and (d)~tetratic ($m=4$) and hexatic ($n=6$).  Note that $n$ is a multiple of $m$ in the first three cases, but not in the fourth case.  We use a lattice model to simulate the combined orientational order in two geometries that require defects:  a disk with strong radial anchoring (which requires a total topological charge of +1), and the surface of a sphere (which requires a total topological charge of +2).

The main results of this study are as follows.  If the coupling between $m$-atic and $n$-atic order is weak, then the system forms a network of point defects connected by domain walls.  The relationship between $m$-atic and $n$-atic order is fixed within each domain, and it shifts across each wall.  As the coupling increases, the domain walls become sharper and shorter.  The behavior for larger coupling depends on whether $n$ is a multiple of $m$.  If so, the system forms clusters with one defect in the $m$-atic order and $n/m$ defects in the $n$-atic order; each cluster has a topological charge of $1/m$.  Eventually, for very strong coupling, the $n$-atic defects merge into the $m$-atic defect, so that the cluster becomes a single defect with a stretched core.  By comparison, if $n$ is not a multiple of $m$, then the system remains as a global network of domains and domain walls.

\section{Phases with a single type of orientational order}

To develop the model, we first simulate phases with a single type of $m$-atic orientational order.  We consider two geometries that require the presence of defects.  The first geometry is a flat disk with strong radial anchoring on the edge.  This anchoring condition requires the total topological charge of +1 inside the disk.  The second geometry is a spherical surface, with orientational order in the local tangent plane.  On a sphere, topology requires a total topological charge of +2.

For the flat disk, we set up a lattice model by constructing a finite-element mesh, using the mesh algorithm pygmsh, a Python interface to Gmsh.  The circular disk geometry is defined parametrically by the radius $R=1$ of the disk, and a triangular discretization is generated using Gmsh’s built-in meshing algorithms.  The mesh size parameter $a=0.1$ controls the element size of the mesh.  On each site $i$ of the mesh, we define an orientation by the unit vector $\hat{\bm{s}}_i=(\cos\theta_i,\sin\theta_i)$.  For a single type of $m$-atic orientational order, we use the Hamiltonian
\begin{equation}
H_m=-J_m\sum_{\langle i,j\rangle}\cos m(\theta_i-\theta_j).
\label{Hmflat}
\end{equation}
In this expression, the sum is over nearest-neighbor sites $i$ and $j$ in the mesh, and the coefficient $J_m>0$ represents the strength of the $m$-atic interaction.  On edge sites, the orientation is constrained to point radially outward, with $\theta_i=\tan^{-1}(y_i/x_i)$.

We perform Monte Carlo simulations of this lattice model.  In this procedure, we begin at a high temperature $T$, and gradually reduce $T$ to zero.  At each step, we make a trial change of $\theta_i$ at one site, and accept or reject this change following the Metropolis algorithm.  The trial change may be either a small rotation or a large rotation of $2\pi/m$ (except for polar order $m=1$, where the large rotation is $\pi$), so that the orientation is free to explore all of the minima of the interaction potential.     

\newlength{\mywidth}
\setlength{\mywidth}{229pt}
\begin{figure}
\centering
a\includegraphics[width=0.48\mywidth]{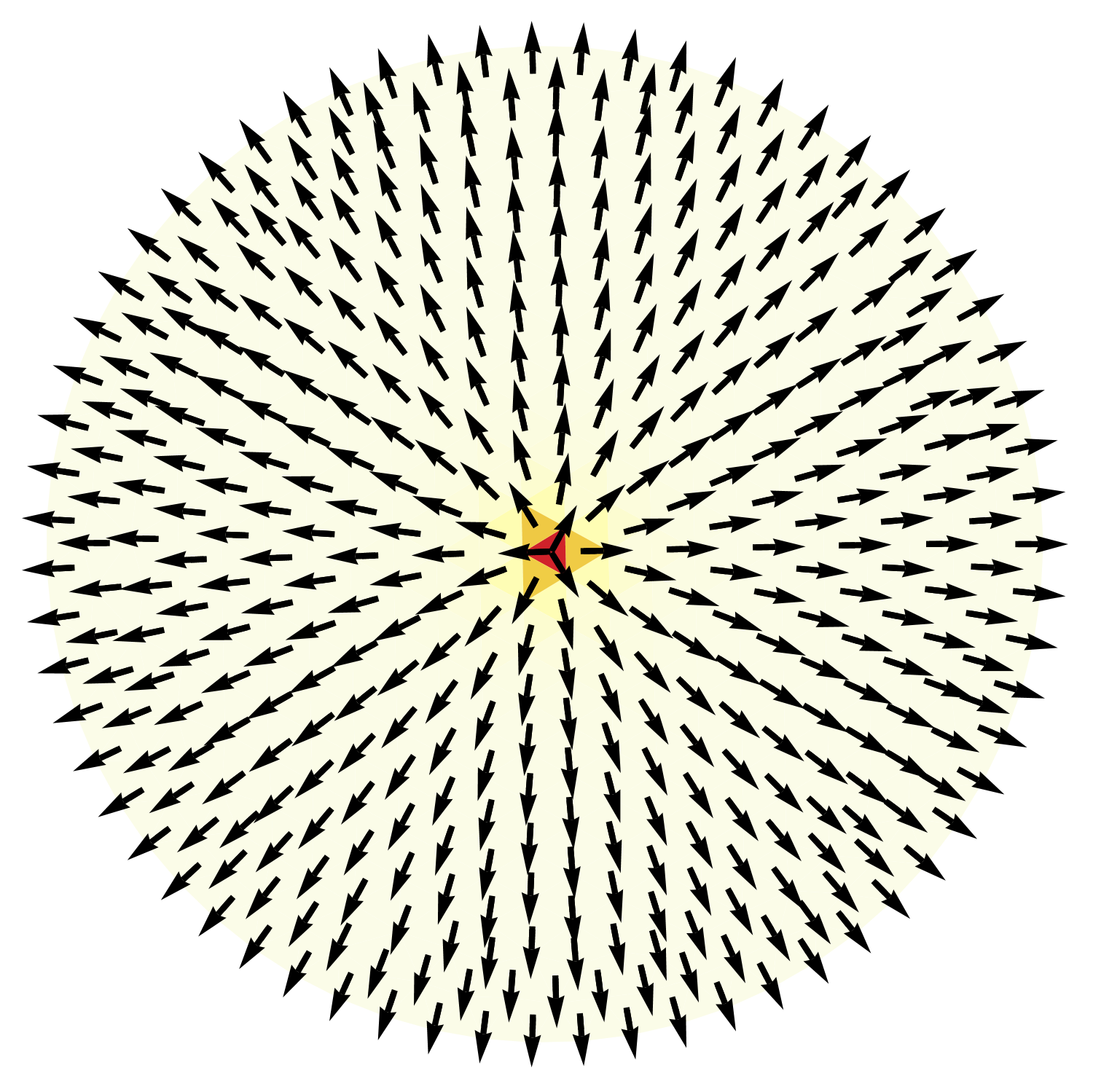}%
b\includegraphics[width=0.48\mywidth]{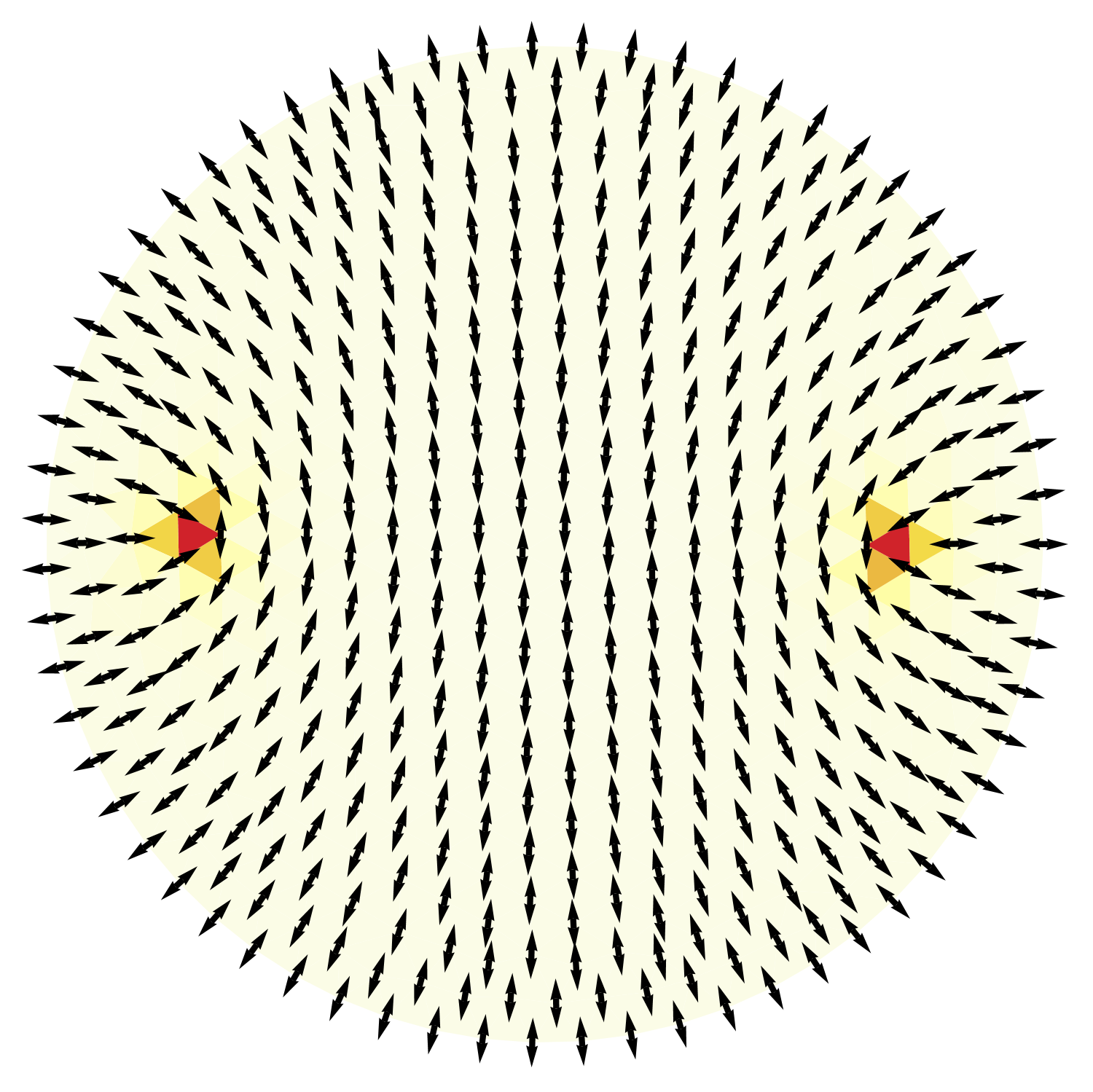}\\
c\includegraphics[width=0.48\mywidth]{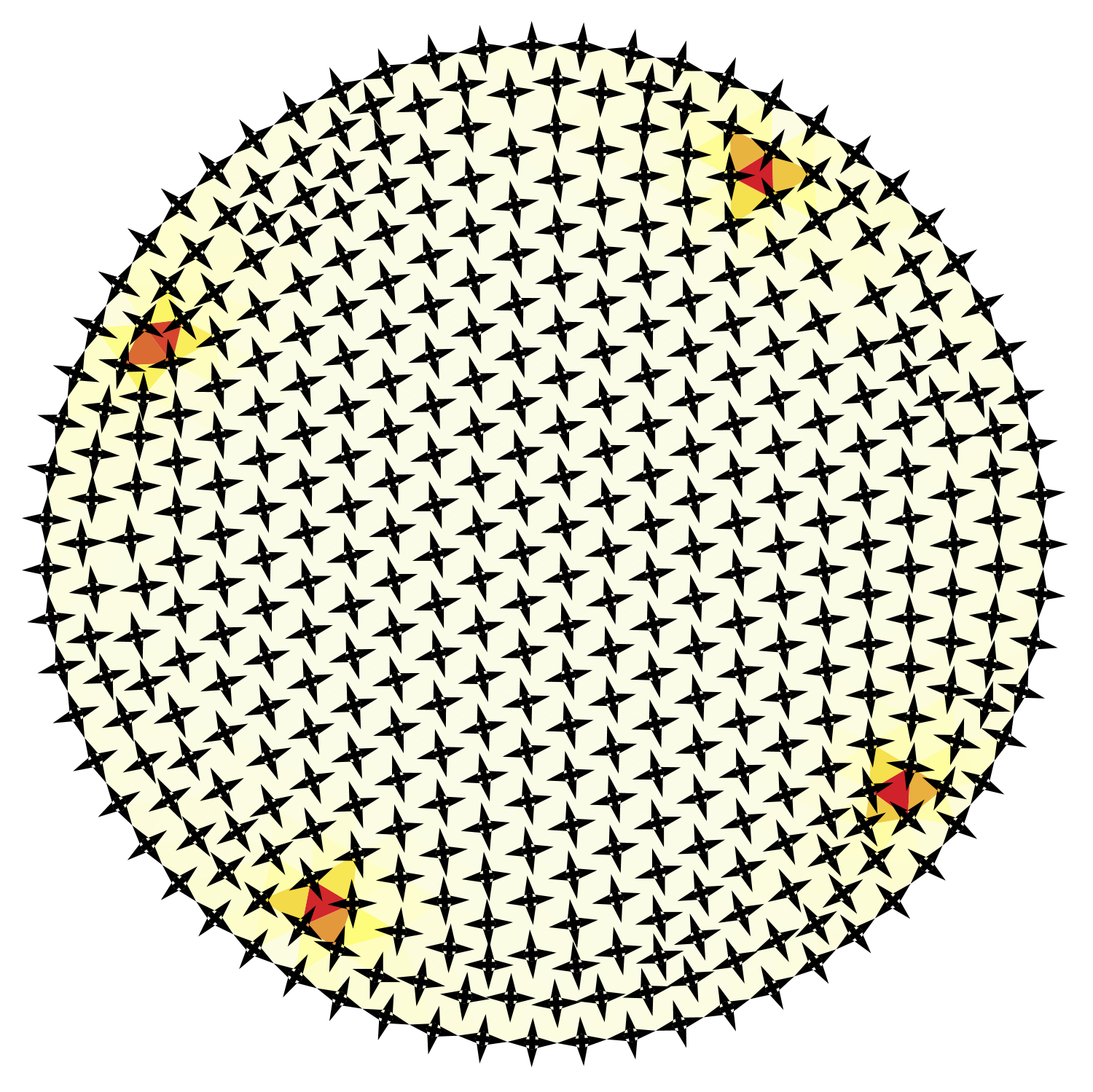}%
d\includegraphics[width=0.48\mywidth]{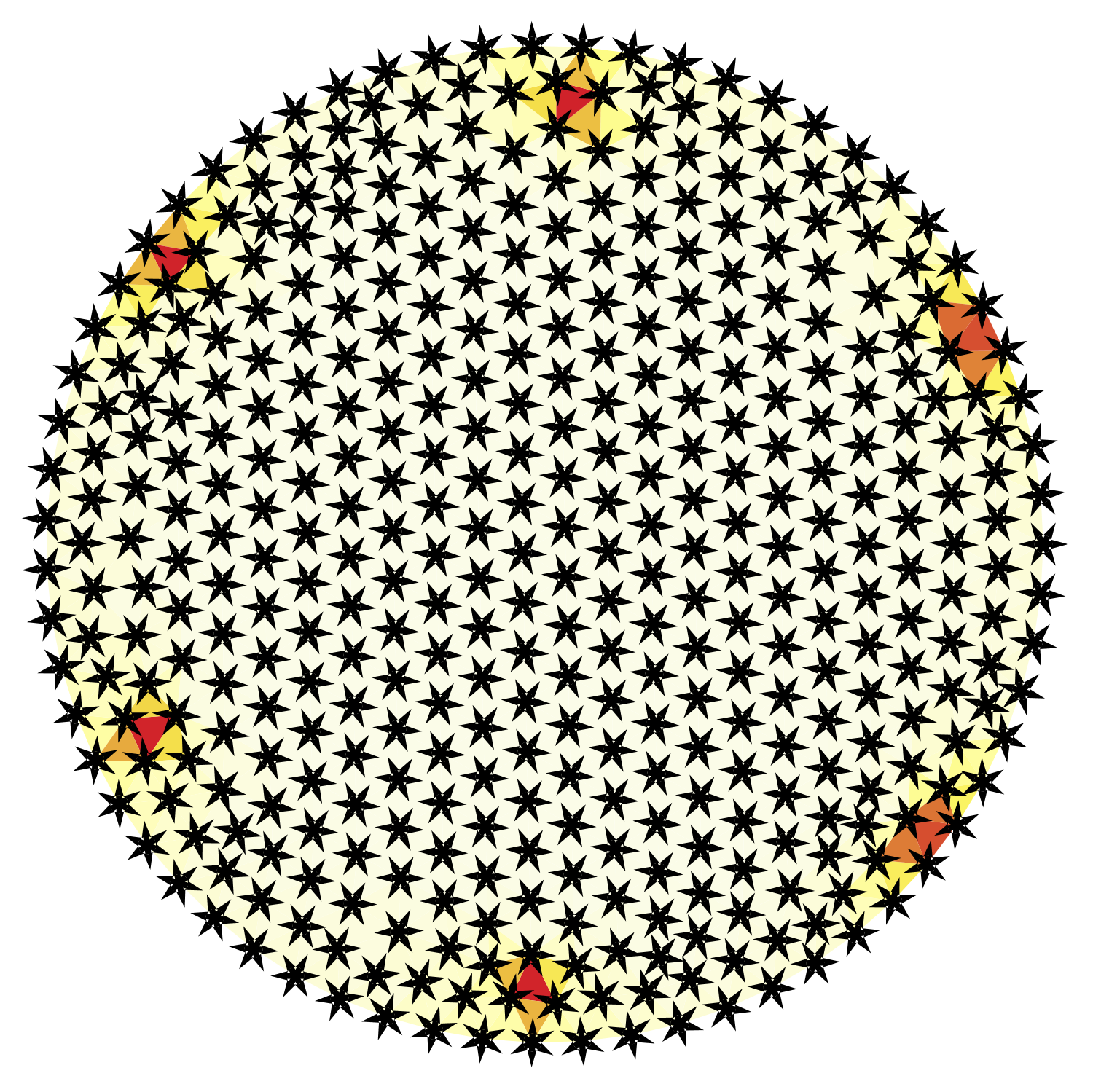}
\caption{Disk with a single $m$-atic type of orientational order.  (a)~Polar.  (b)~Nematic.  (c)~Tetratic.  (d)~Hexatic.  At each site of the mesh, the $m$ arrows indicate the local orientational order with $m$-fold symmetry.  On each triangular cell, the color (yellow to red) indicates the local energy density (average of the three bond energies around the cell).}
\label{fig:disksingleorder}
\end{figure}

Figure~\ref{fig:disksingleorder} shows results for polar ($m=1$), nematic ($m=2$), tetratic ($m=4$), and hexatic ($m=6$) interactions.  In this figure, we visualize the results by drawing $m$ arrows at each mesh site, corresponding to the orientations $\theta_i$, $\theta_i+2\pi/m$, \dots, ${\theta_i+2\pi(m-1)/m}$.  This visualization is appropriate because all $m$ of those arrows represent equivalent directions, all with the same energy, showing the $m$-fold symmetry.  We also color each triangular cell of the mesh to indicate the local energy density.  

In each case, the simulated annealing algorithm reaches an energy minimum, in which the system has well-defined $m$-atic order everywhere except at point defects.  As expected, the disk has $m$ defects, each with topological charge $+1/m$, for a total topological charge of $+1$.  Around each defect, the orientation of the arrows rotates through an angle of $2\pi/m$.  The defect cores have high energy density, as indicated by the red color in the figure. 

For simulations on a sphere, we follow a similar procedure, but we must be careful with the coordinate system.  The 2D spherical coordinate system has singularities at the north and south poles, and these singularities might bias the defect structure on the sphere.  To avoid any such bias, we explicitly work in 3D Cartesian coordinates, generalizing the approach from previous studies of polar and nematic order on a sphere~\cite{Shin2008,Selinger2011}.  We construct a finite-element mesh on a spherical surface with radius~1, again using the mesh algorithm in the Python library, trimesh package.  The function trimesh.creation.icosphere() generates a sphere using the icosphere method, which starts from an icosahedron, a regular polyhedron with 20 faces, and iteratively subdivides the triangles to create a more spherical shape.  On each site $i$ of the mesh, with position $\hat{\bm{r}}_i=(x_i,y_i,z_i)$, we define an orientation by the unit vector $\hat{\bm{s}}_i=(s_{ix},s_{iy},s_{iz})$.  We require that all of the vectors are normalized to $|\hat{\bm{s}}_i|=1$, and maintain this normalization at each step of the simulation.  To force these vectors into the local tangent plane, we include a term in the Hamiltonian of the form
\begin{equation}
H_\text{normal}=C\sum_i(\hat{\bm{r}}_i\cdot\hat{\bm{s}}_i)^2,
\label{Hnormal}
\end{equation}
with a large coefficient $C>0$.  To express the $m$-atic interaction in terms of the 3D vectors, we define $\theta_{ij}$ as the angle between $\hat{\bm{s}}_i$ and $\hat{\bm{s}}_j$, so that $\cos\theta_{ij}=\hat{\bm{s}}_i\cdot\hat{\bm{s}}_j$.  The $m$-atic interaction then becomes
\begin{equation}
H'_m=-J_m\sum_{\langle i,j\rangle}\cos m\theta_{ij}.
\label{Hmsphere}
\end{equation}
Using trigonometric identities, it can be transformed into
\begin{align}
H'_1&=-J_1\sum_{\langle i,j\rangle}\hat{\bm{s}}_i\cdot\hat{\bm{s}}_j,\qquad
H'_2=-J_2\sum_{\langle i,j\rangle}\left[2(\hat{\bm{s}}_i\cdot\hat{\bm{s}}_j)^2-1\right],\nonumber\\
H'_4&=-J_4\sum_{\langle i,j\rangle}\left[8(\hat{\bm{s}}_i\cdot\hat{\bm{s}}_j)^4-8(\hat{\bm{s}}_i\cdot\hat{\bm{s}}_j)^2+1\right],\nonumber\\
H'_6&=-J_6\sum_{\langle i,j\rangle}\left[32(\hat{\bm{s}}_i\cdot\hat{\bm{s}}_j)^6-48(\hat{\bm{s}}_i\cdot\hat{\bm{s}}_j)^4+18(\hat{\bm{s}}_i\cdot\hat{\bm{s}}_j)^2-1\right].
\label{Hmsphereexamples}
\end{align}

We perform Monte Carlo simulations of the total lattice Hamiltonian $H_\text{normal}+H'_m$, beginning at high temperature $T$ and gradually reducing $T$ to zero.  At each step, we make a trial change of $\hat{\bm{s}}_i$ at one site, and accept or reject this change following the Metropolis algorithm.  The trial change may be either a small rotation about a random axis or a large rotation of $2\pi/m$ about $\hat{\bm{r}}_i$ (except for polar order $m=1$, where the large rotation is $\pi$), so that the orientation is free to explore all of the minima of Hamiltonian.  

\begin{figure}
\centering
a\includegraphics[height=0.48\mywidth]{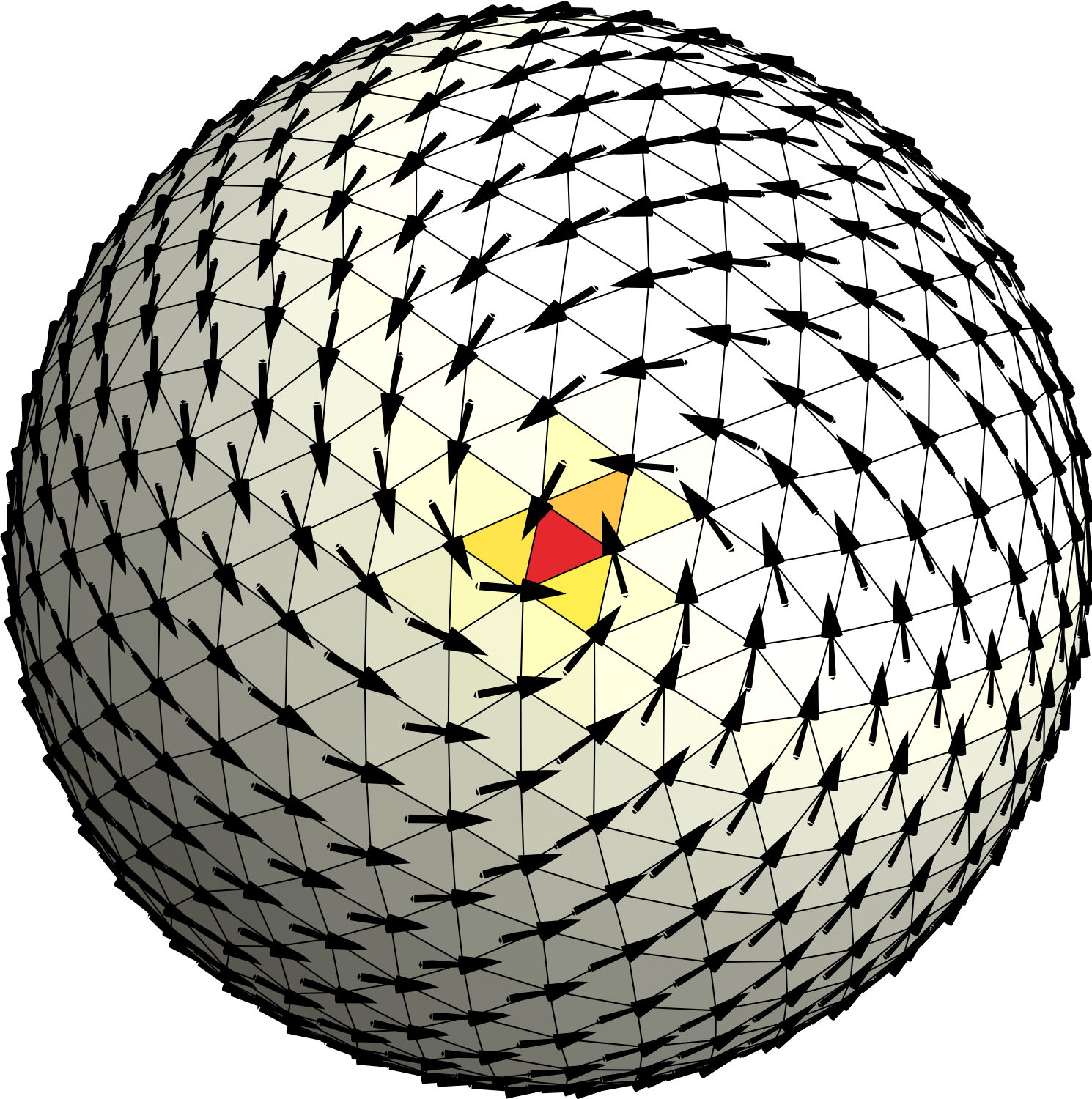}
b\includegraphics[height=0.48\mywidth]{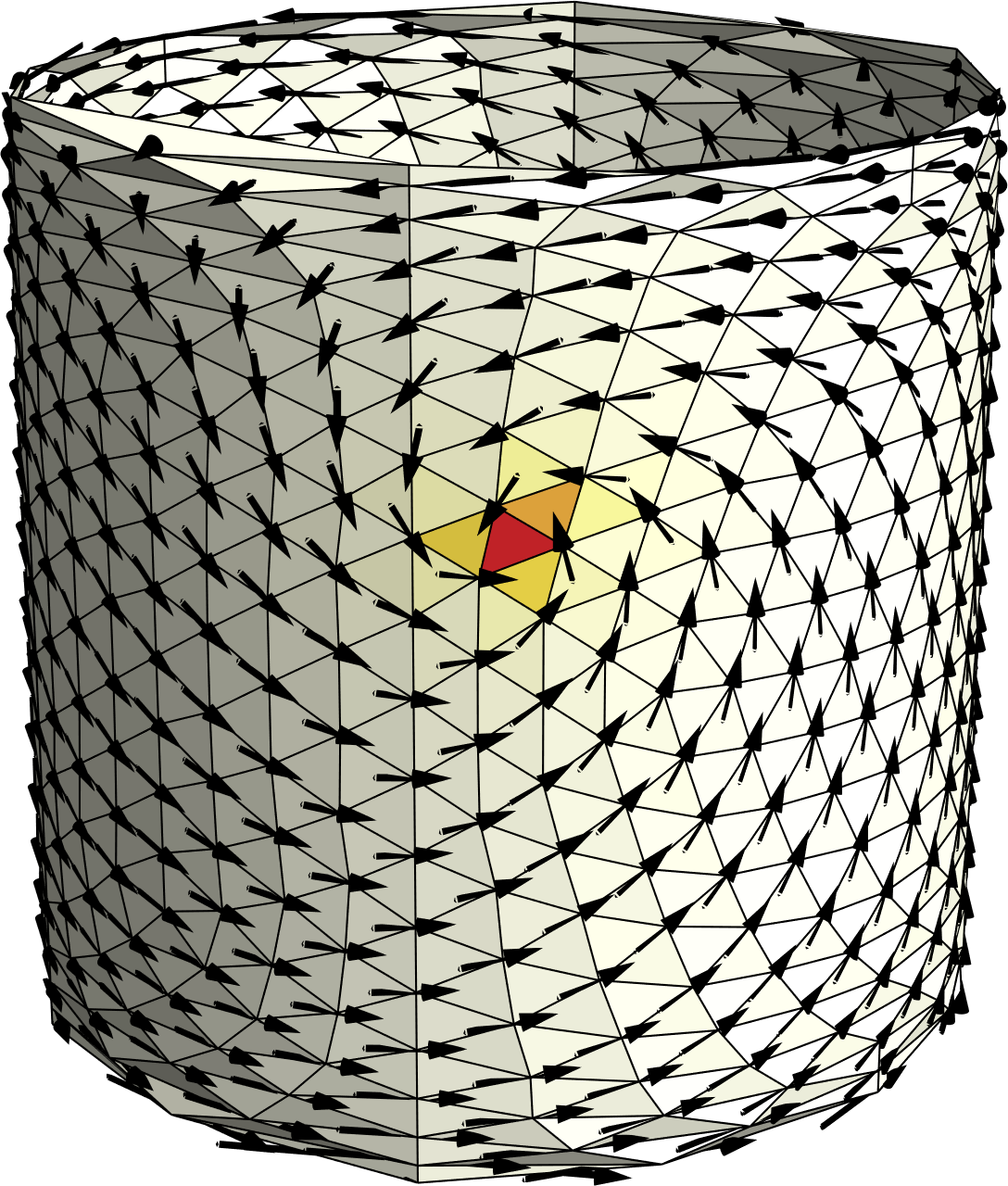}\\
\smallskip
c~\includegraphics[width=.97\mywidth]{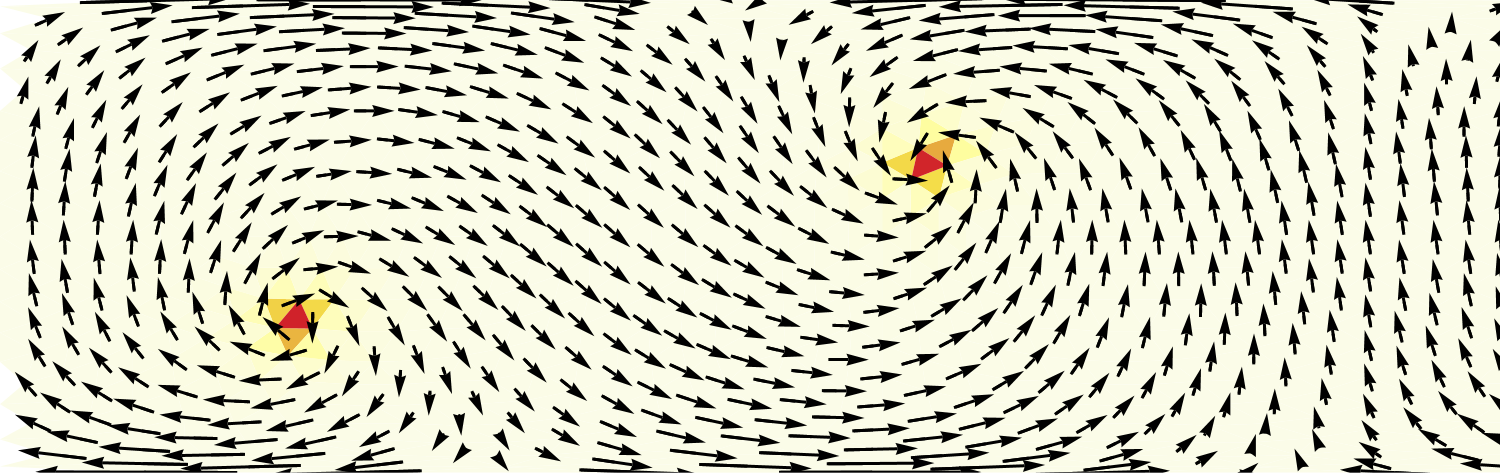}\\
\smallskip
d~\includegraphics[width=.97\mywidth]{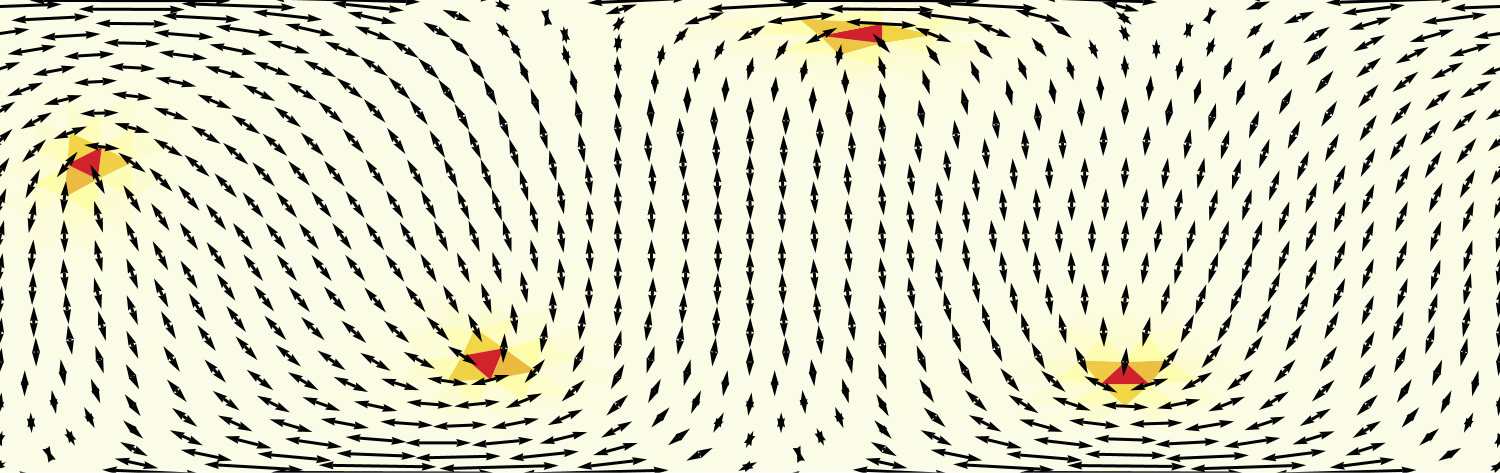}\\
\smallskip
e~\includegraphics[width=.97\mywidth]{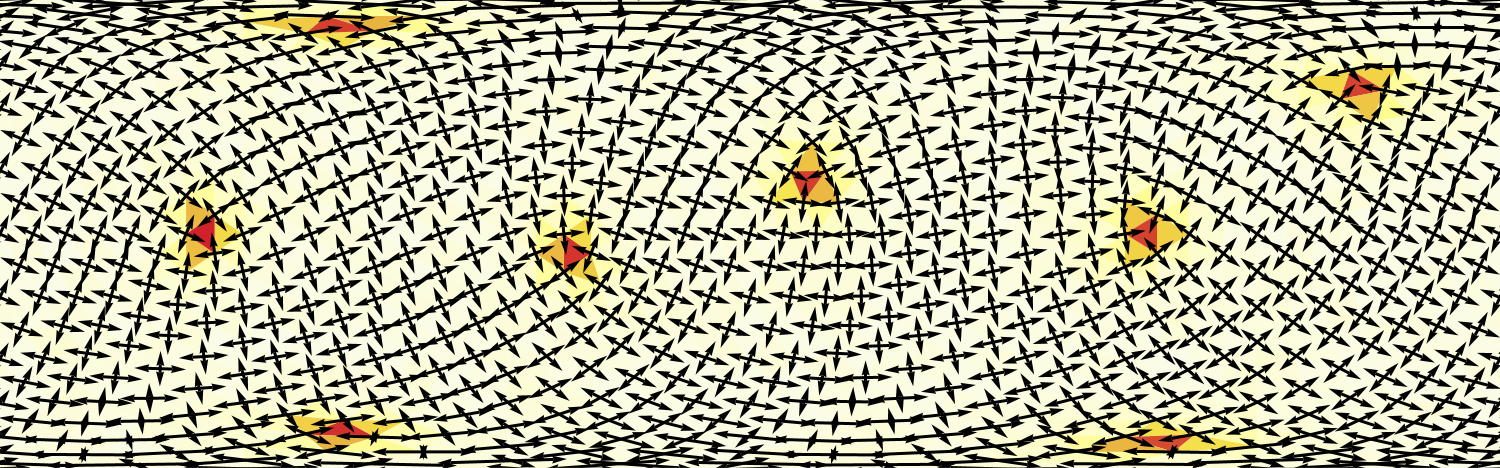}\\
\smallskip
f~\includegraphics[width=.97\mywidth]{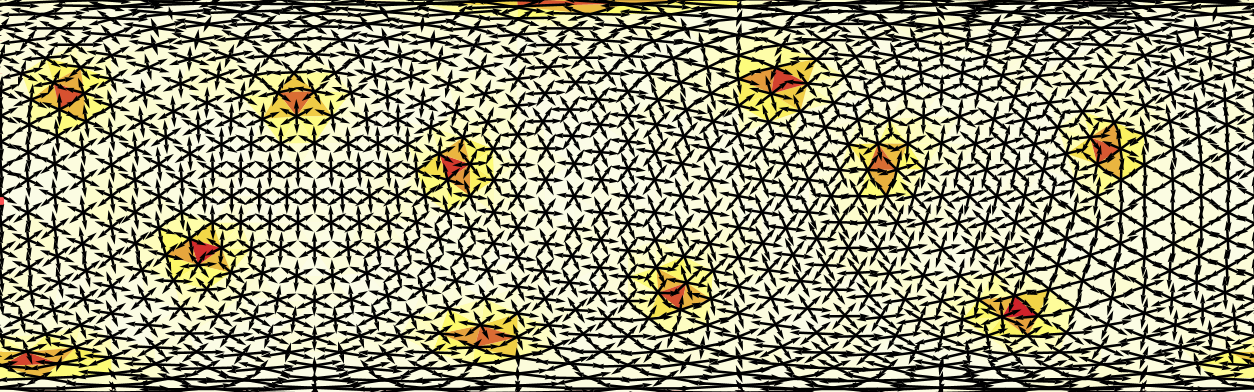}
\caption{Sphere with a single $m$-atic type of orientational order.  (a)~Polar, shown on a sphere, only one defect visible.  (b)~Polar, projected onto a cylinder, only one defect visible.  (c)~Polar, Mercator projection onto flat plane, both defects visible.  (d)~Nematic, Mercator projection, all four defects visible.  (e)~Tetratic, Mercator projection, all eight defects visible.  (f)~Hexatic, Mercator projection, all twelve defects visible.  As in Fig.~\ref{fig:disksingleorder}, the arrows indicate the local orientational order with $m$-fold symmetry, and the color indicates the local energy density.}
\label{fig:spheresingleorder}
\end{figure}

Figure~\ref{fig:spheresingleorder}(a) shows results for polar order, visualized on the sphere.  On each mesh site, the arrow represents the unit vector $\hat{\bm{s}}_i$.  On each triangular cell of the mesh, the color indicates the local energy density.  The polar order is well-defined everywhere except at two point defects, each with topological charge +1, for a total topological charge of +2.  One of the defects is clearly visible in the figure, as shown by the red color for high energy density.  The other defect cannot be seen because it is on the back of the sphere.

For a more global view of the entire defect configuration, we must modify our visualization.  First, we project the sphere onto a cylinder, as shown in Fig.~\ref{fig:spheresingleorder}(b).  Next, we unroll the cylinder into the flat plane, as in Fig.~\ref{fig:spheresingleorder}(c).  This construction is the Mercator projection, which is well-known in map-making.  The same construction was recently used to visualize colloidal crystals on a sphere~\cite{Jones2025}.  For our current study, the main advantage of the Mercator projection is that it shows the orientational order on the entire sphere.  In the system with polar order, it shows both of the defects.  For that reason, we will use the Mercator projection of the sphere for the rest of this paper.  Of course, we must recognize that the Mercator projection distorts the structure near the north and south poles.

Figures~\ref{fig:spheresingleorder}(d--f) show similar results for nematic ($m=2$), tetratic ($m=4$), and hexatic ($m=6$) orientational order.  In each case, we draw $m$ arrows at each mesh site, corresponding to the orientation  $\hat{\bm{s}}_i$ rotated through angles $0$, $2\pi/m$, \dots, $2\pi(m-1)/m$ about the local normal vector $\hat{\bm{r}}_i$.  This visualization is appropriate because all $m$ of those arrows represent equivalent directions in the local tangent plane.  We can see that the sphere has well-defined $m$-atic order everywhere except at point defects.  There are $2m$ defects, each with topological charge $+1/m$, for a total topological charge of $+2$.

\section{Phases with two types of orientational order}

\begin{figure}
\centering
a\includegraphics[width=.97\mywidth]{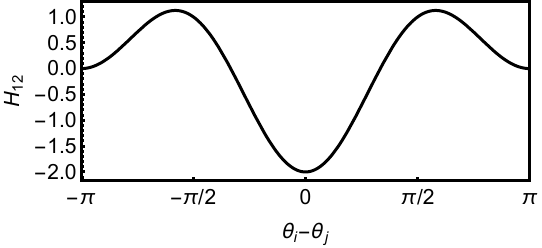}
b\includegraphics[width=.97\mywidth]{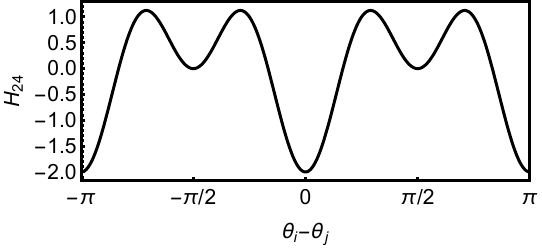}
c\includegraphics[width=.97\mywidth]{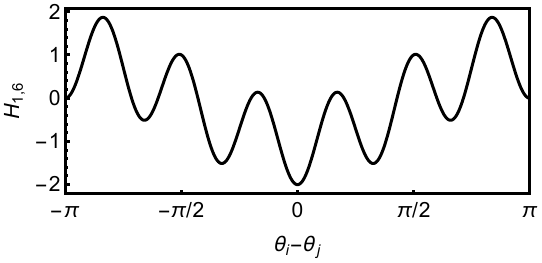}
d\includegraphics[width=.97\mywidth]{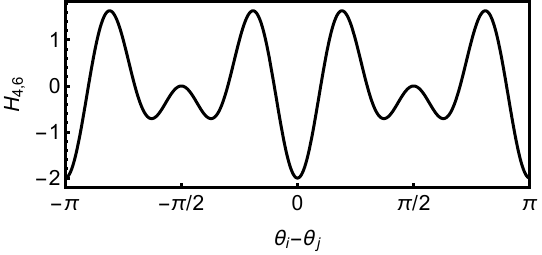}
\caption{Plots of the potential energy $H_{m,n}$ as a function of the relative orientation of neighboring sites, for the four examples considered in this paper.  (a)~Polar and nematic.  (b)~Nematic and tetratic.  (c)~Polar and hexatic.  (d)~Tetratic and hexatic.  In each case, the coefficients are $J_m=J_n=1$.}
\label{fig:potentialplots}
\end{figure}

We now generalize the model to simulate phases with two types of orientational order, $m$-atic and $n$-atic, which interact with each other.  In this generalization, we combine the Hamiltonian functions from the previous section, corresponding to $m$-atic and $n$-atic order.  For a flat disk, we add the functions $H_m$ and $H_n$ from Eq.~(\ref{Hmflat}), to obtain
\begin{equation}
H_{m,n}=-\sum_{\langle i,j\rangle}[J_m\cos m(\theta_i-\theta_j)+J_n\cos n(\theta_i-\theta_j)].
\label{Hmnflat}
\end{equation}
Similarly, for a sphere, we add the functions of Eqs.(\ref{Hnormal}--\ref{Hmsphereexamples}) to obtain $H_\text{total}=H_\text{normal}+H'_m+H'_n$.  Figure~\ref{fig:potentialplots} shows plots of this potential energy for the four examples considered in this paper---polar (${m=1}$) and nematic ($n=2$), nematic ($m=2$) and tetratic ($n=4$), polar ($m=1$) and hexatic ($n=6$), and tetratic ($m=4$) and hexatic ($n=6$)---always with $J_m=J_n=1$.  In the first three cases, where $n$ is a multiple of $m$, the potential energy has absolute minima when the neighboring orientations are aligned with each other modulo $2\pi/m$, and further local minima when they are aligned with each other modulo $2\pi/n$.  In the fourth case, where $n$ is not a multiple of $m$, the shape of the potential is more complex, as will be discussed in Section~\ref{sec:tetratichexatic} below.

\subsection{Polar and nematic}

As a first example, we begin with the combination of polar and nematic order.  This is the simplest case, which was originally considered by Lee and Grinstein~\cite{Lee1985}.  It is also the case that is most relevant for ferroelectric nematic liquid crystals.

\begin{figure}
\centering
a\includegraphics[width=.48\mywidth]{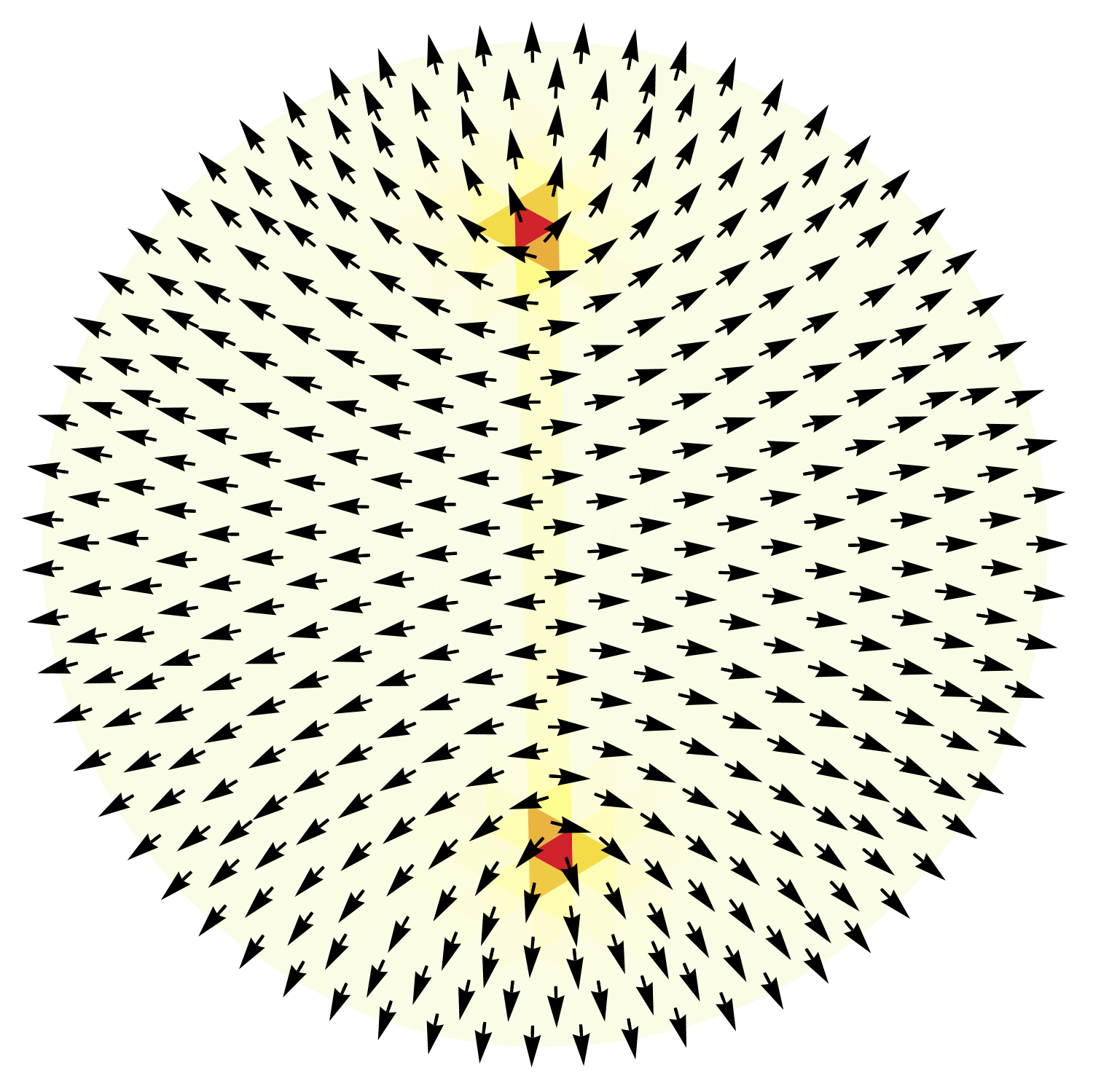}%
b\includegraphics[width=.48\mywidth]{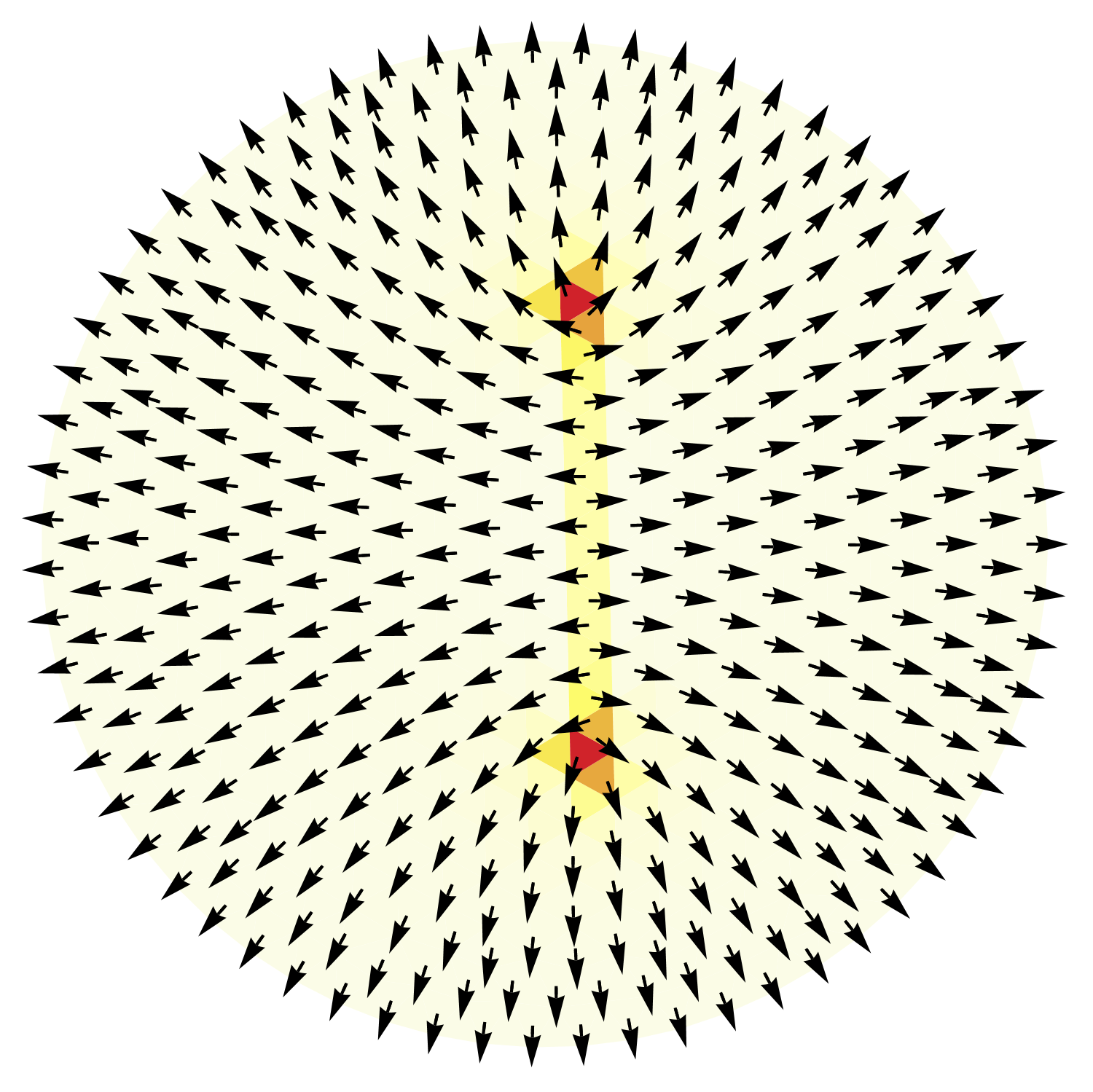}\\
c\includegraphics[width=.48\mywidth]{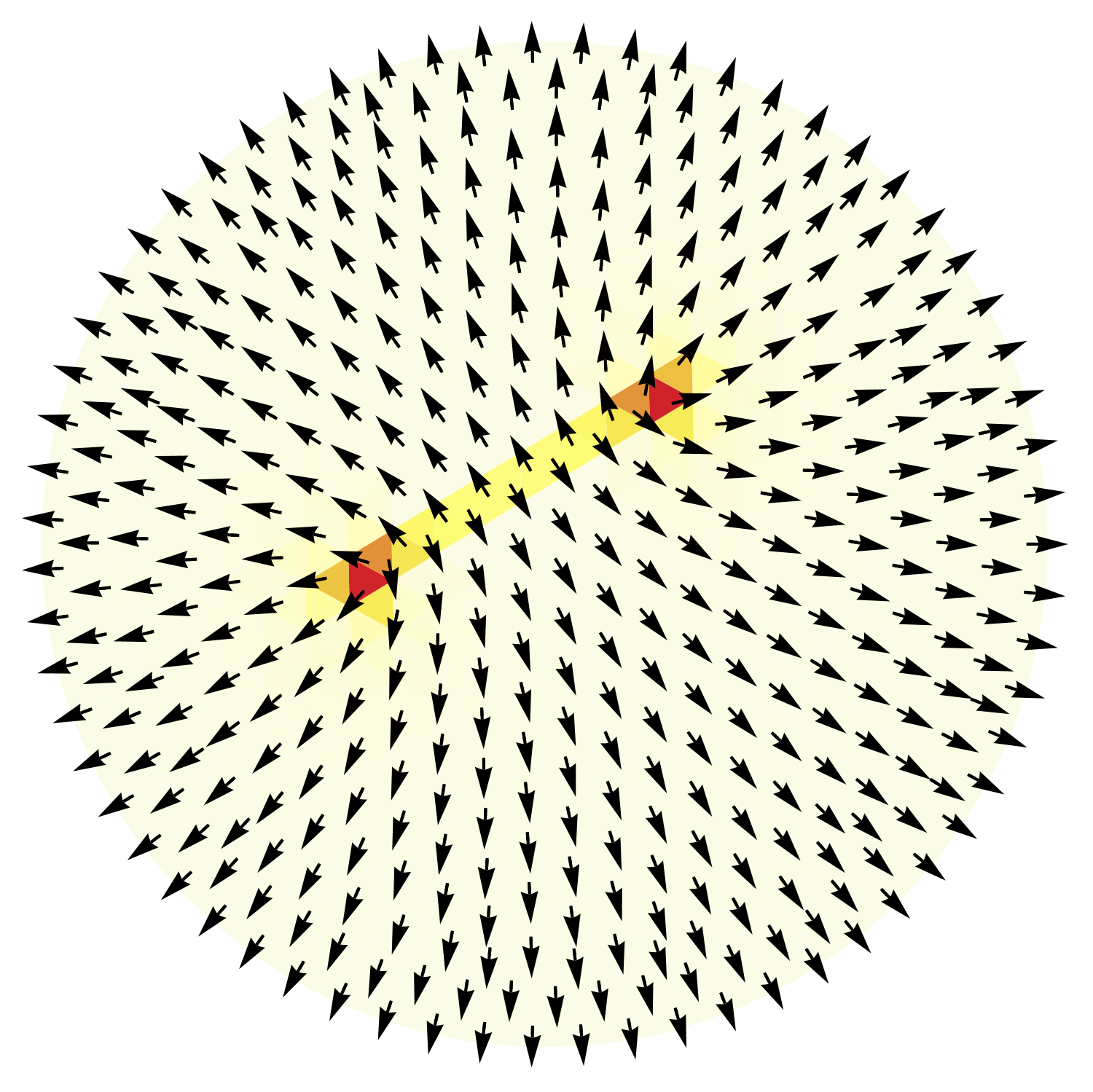}%
d\includegraphics[width=.48\mywidth]{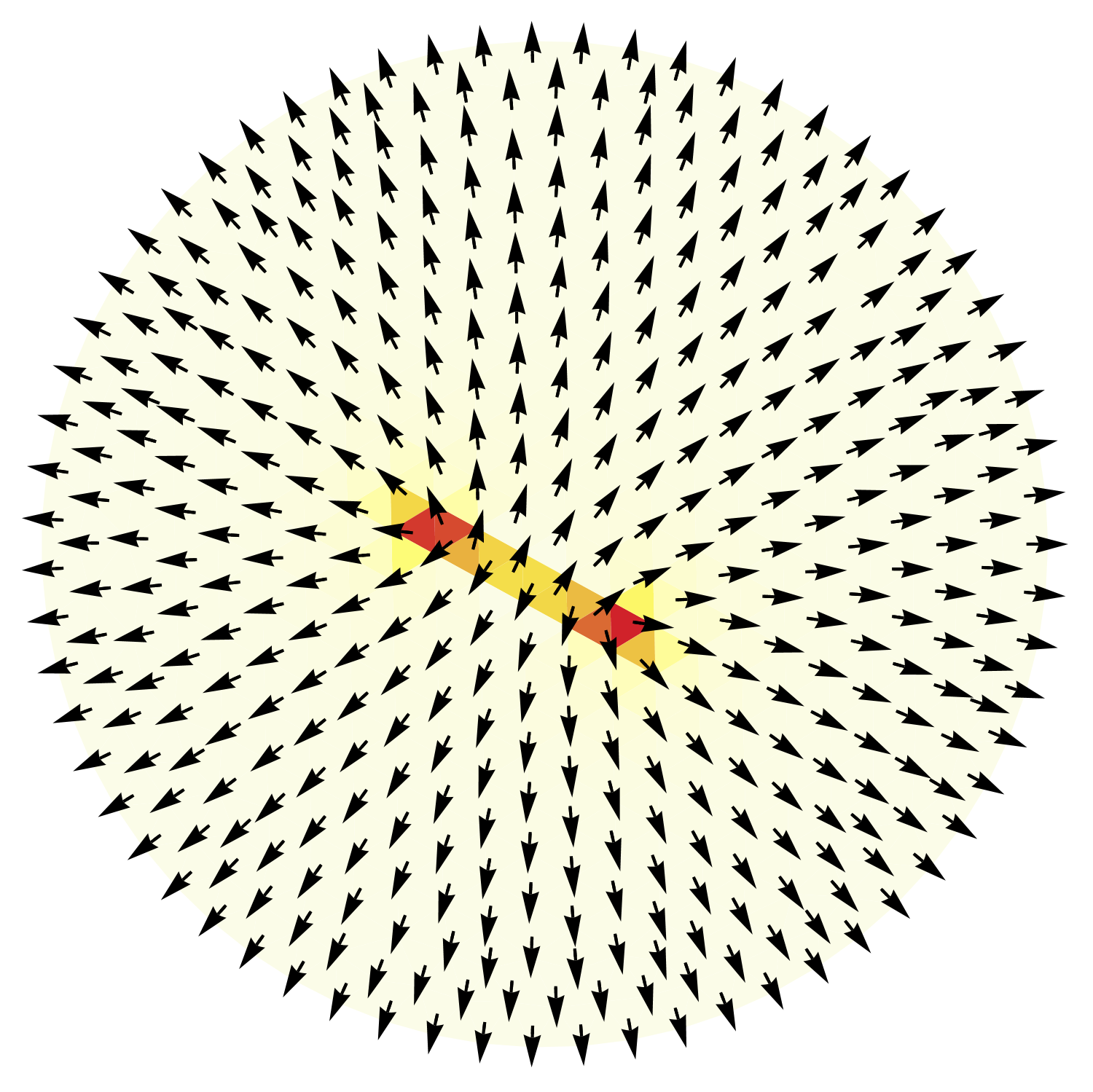}\\
\smallskip
e~\includegraphics[width=.97\mywidth]{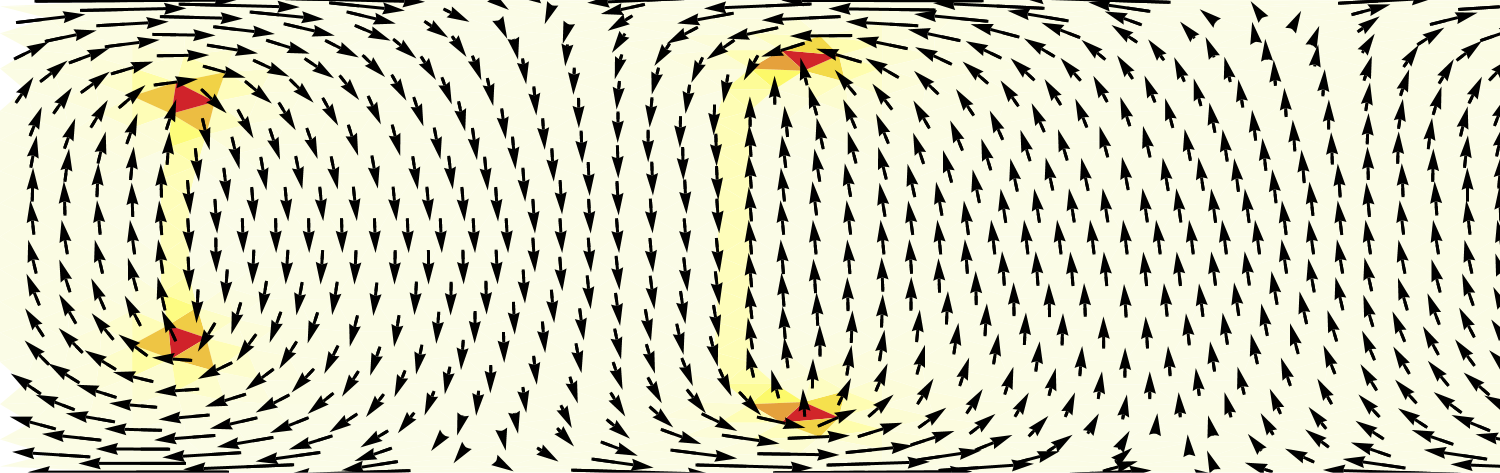}\\
\smallskip
f~\includegraphics[width=.97\mywidth]{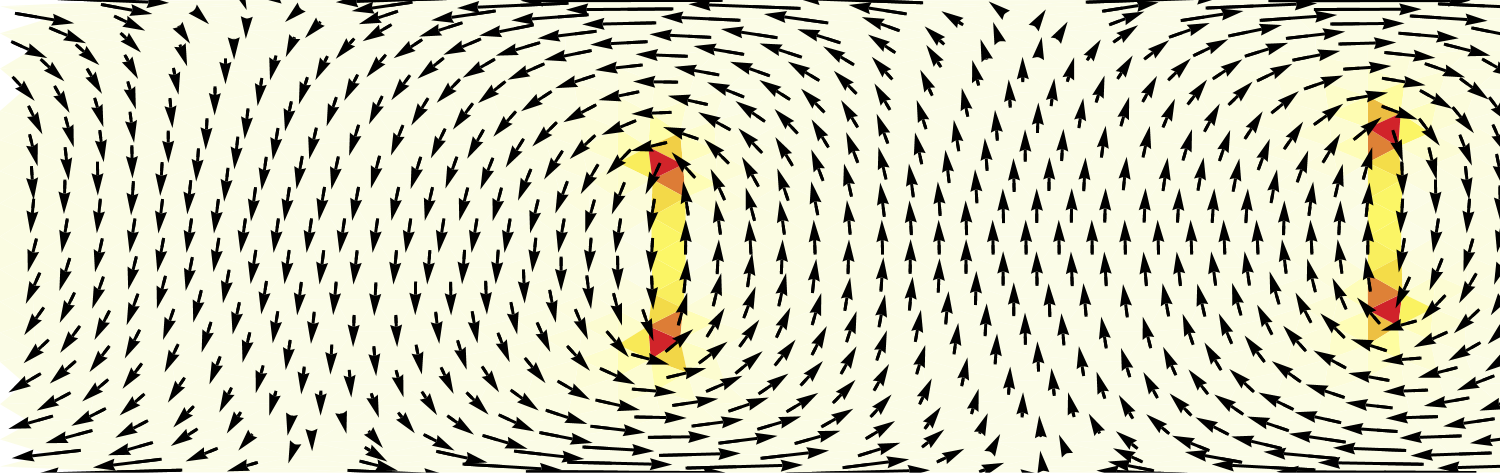}\\
\smallskip
g~\includegraphics[width=.97\mywidth]{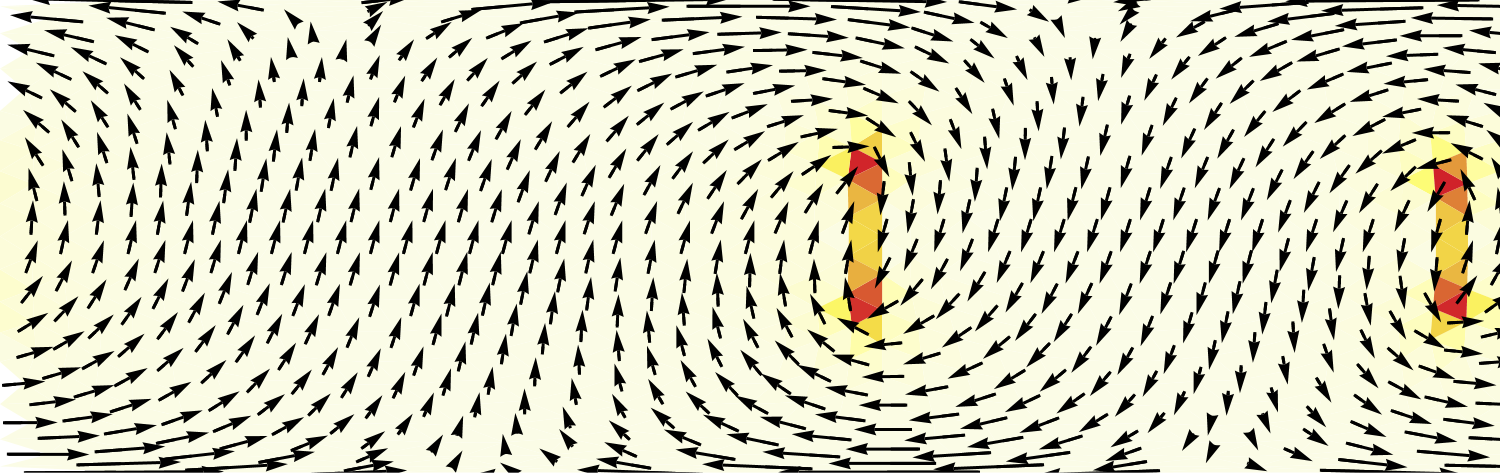}\\
\smallskip
h~\includegraphics[width=.97\mywidth]{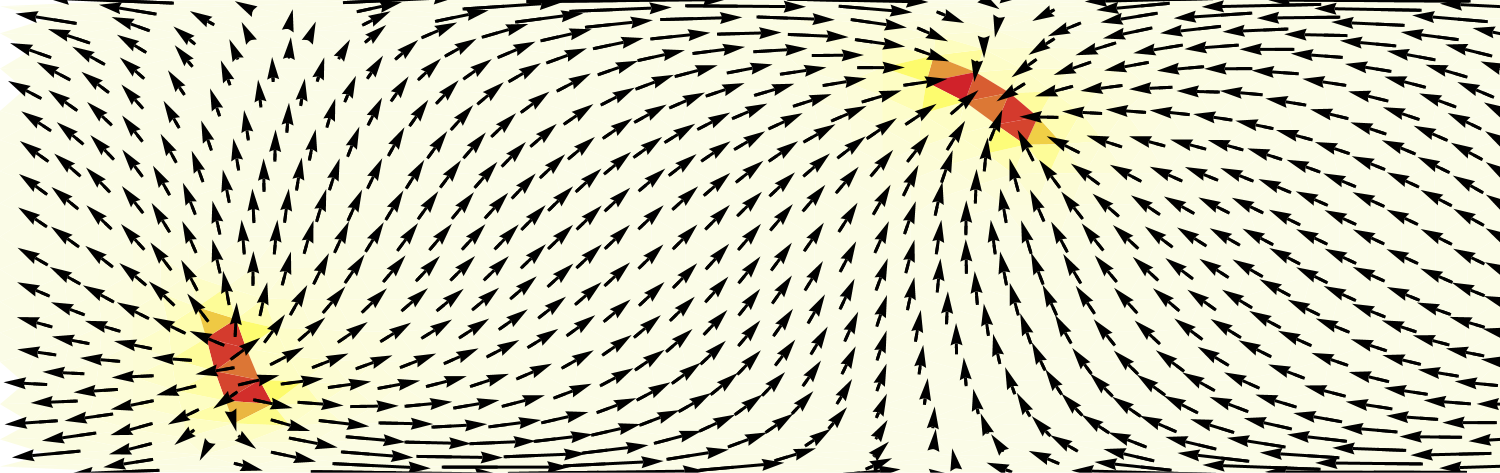}
\caption{Phase with combined polar and nematic order.  (a--d) Simulation on disk with $J_2=2$ and $J_1=0.2$, $0.5$, $0.8$, and $1.5$, respectively.  (e-h) Mercator projection of sphere with $C=5$, $J_2=1$, and $J_1=0.2$, $0.5$, $0.8$, and $1.5$.}
\label{fig:polarnematic}
\end{figure}

Figure~\ref{fig:polarnematic}(a--d) shows a series of simulations of the Hamiltonian $H_{1,2}$ on a disk with radial boundary conditions.  The nematic interaction strength is fixed at $J_2=2$, and the polar interaction strength increases over the range $J_1=0.2$, $0.5$, $0.8$, and $1.5$.  The arrows indicate the local orientational order.  In most of the disk, away from any defects, the phase has polar order, as shown by alignment of neighboring arrows.  It also has nematic order, shown by the alignment of neighboring arrow shafts, neglecting the arrow heads.  In those respects, it is similar to the polar phase shown in Fig.~\ref{fig:disksingleorder}(a).  The difference occurs in the defects.

In Fig.~\ref{fig:polarnematic}(a), where the polar coefficient $J_1$ is smallest, the phase has two point defects with high energy density, indicated by the red color.  These two points are topological defects in the nematic order, with topological charge of $+1/2$ each.  They are connected by a domain wall with moderate energy density, indicated by yellow.  Across the domain wall, the polar order reverses, while the nematic order remains unchanged.  In Fig.~\ref{fig:polarnematic}(b--d), as the polar coefficient $J_1$ increases, the domain wall becomes shorter, and the two point defects become closer together.  In all cases, the entire structure (two point defects connected by a domain wall) can be regarded as a single ``string defect,'' with topological charge of $+1$, as required by the boundary conditions.

Figure~\ref{fig:polarnematic}(e--h) shows analogous simulations on a sphere, illustrated in the Mercator projection.  Here, the normal interaction is fixed at $C=5$, the nematic interaction at $J_2=1$, and the polar interaction increases over the range $J_1=0.2$, $0.5$, $0.8$, and $1.5$.  The results are generally similar to the simulations on the disk.  Once again, the system forms string defects of topological charge $+1$, each consisting of two $+1/2$ nematic defects connected by a domain wall.  The sphere has two string defects, for a total topological charge of $+2$, as required by topology.  As the polar coefficient $J_1$ increases, the domain walls become shorter and the pairs of nematic point defects become closer together.  In the limit of strong polar interaction, each pair of nematic defects effectively merges into a single polar defect with an extended core.

These simulation results are consistent with the theory of string defects, as reviewed in Section 7.2 of the textbook~\cite{Selinger2024}.  The free energy of a string defect has two parts.  First, there is the repulsion between the two nematic defects, which is $2\pi K q_1 q_2\log(R/L)$, where $K$ is the orientational elastic constant, $q_1$ and $q_2$ are the separate topological charges ($q_1=q_2=+1/2$), $R$ is the system size, and $L$ is the distance between the defects.  Second, there is the free energy of the domain wall itself, which is $\epsilon_W L$, where $\epsilon_W$ is the line tension (energy per length).  The total is
\begin{equation}
F_\text{string}=\frac{\pi K}{2}\log\left(\frac{R}{L}\right)+\epsilon_W L.
\end{equation}
Minimizing the total free energy over $L$ gives the equilibrium length
\begin{equation}
L=\frac{\pi K}{2\epsilon_W}.
\end{equation}
Here, the effective line tension $\epsilon_W$ is the difference between the energies for antiparallel and parallel alignment of neighboring arrows, normalized by the typical spacing $a$ between mesh points.  We estimate $\epsilon_W\approx2J_1/a$.  The effective elastic constant $K$ is the second derivative of the potential $H_{1,2}$ at the minimum.  We estimate $K\approx(J_1+4J_2)$.  Hence, the equilibrium length of a string defect should be
\begin{equation}
L\approx\frac{\pi a(J_1+4J_2)}{4J_1}.
\end{equation}
For $J_1=0$, this predicted length is infinite, meaning that the two nematic defects are free to move apart, not connected by a domain wall.  As $J_1$ increases, this predicted length decreases, so that the nematic defects are more tightly bound by the domain wall, and the string becomes shorter.  This trend agrees with the simulations.

\subsection{Nematic and tetratic}

Next, we consider the combination of nematic and tetratic order.  Mathematically, this problem is equivalent to the combination of polar and nematic order, but with the angle $2\theta$ substituted in place of $\theta$.  For that reason, we expect to see string defects, similar to the string defects seen in the previous case.  Hence, the main purpose of the new simulations is visualize defects in nematic and tetratic order, so that they can be compared with other research.

\begin{figure}
\centering
a\includegraphics[width=.48\mywidth]{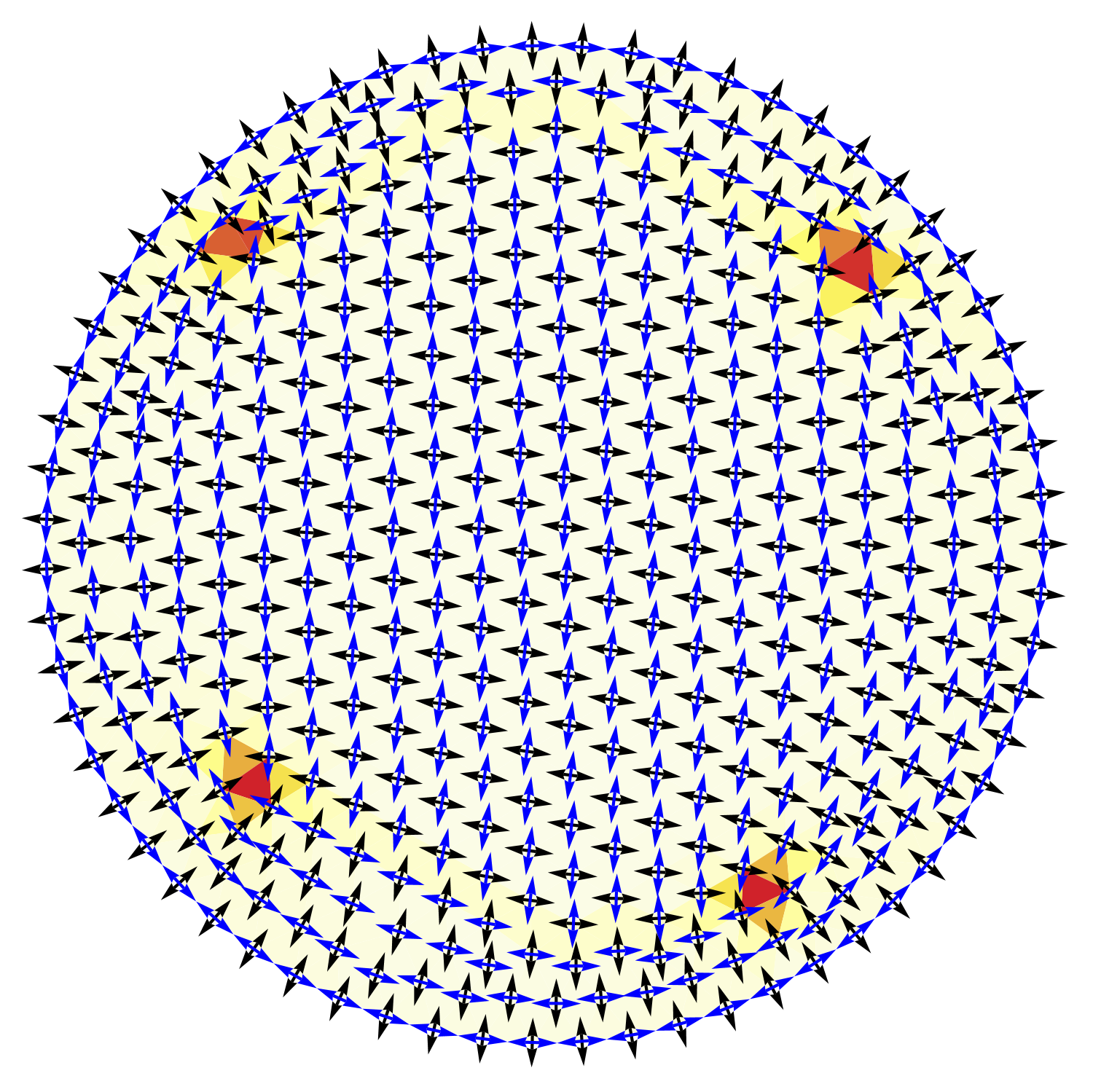}%
b\includegraphics[width=.48\mywidth]{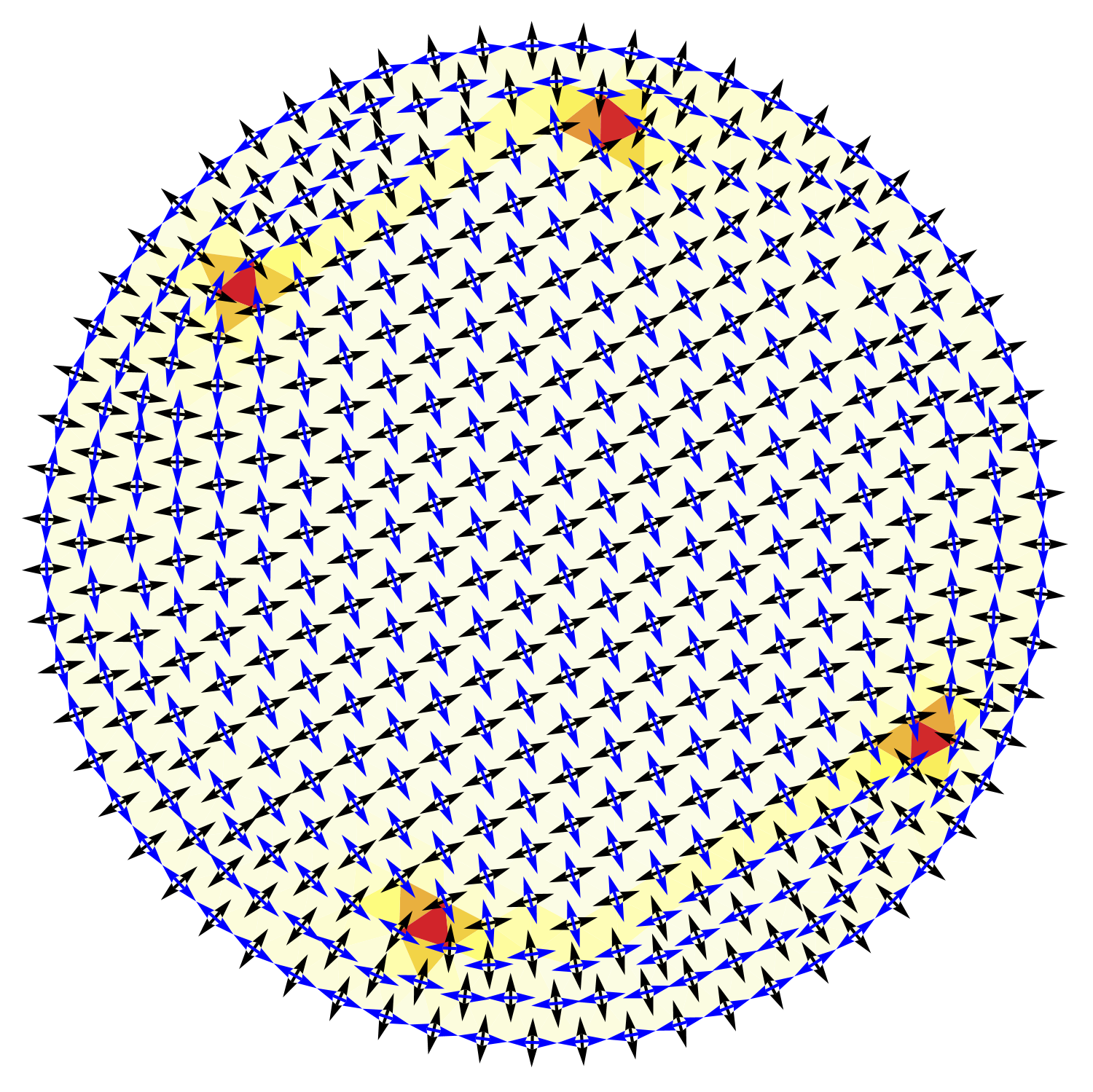}\\
c\includegraphics[width=.48\mywidth]{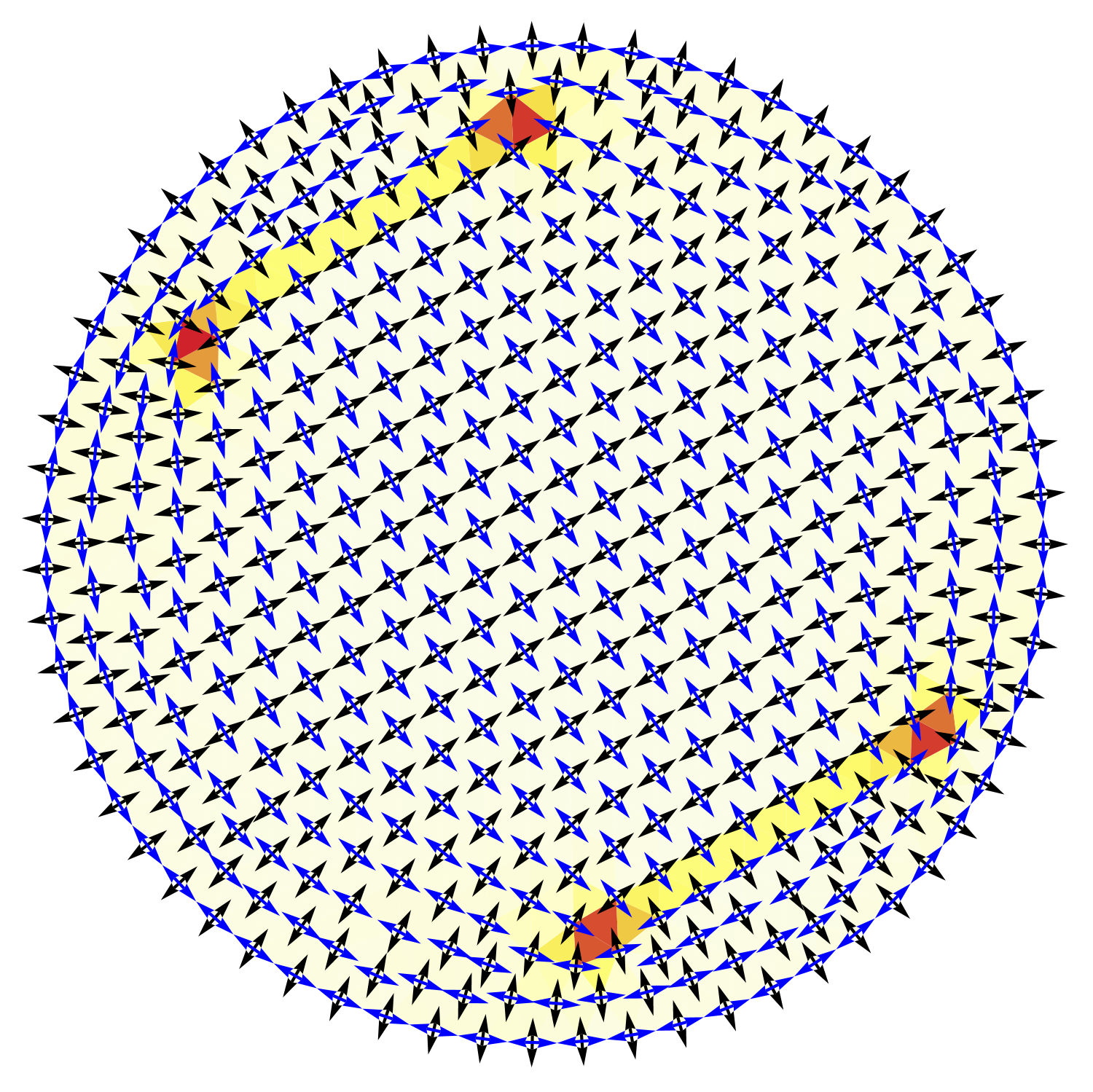}%
d\includegraphics[width=.48\mywidth]{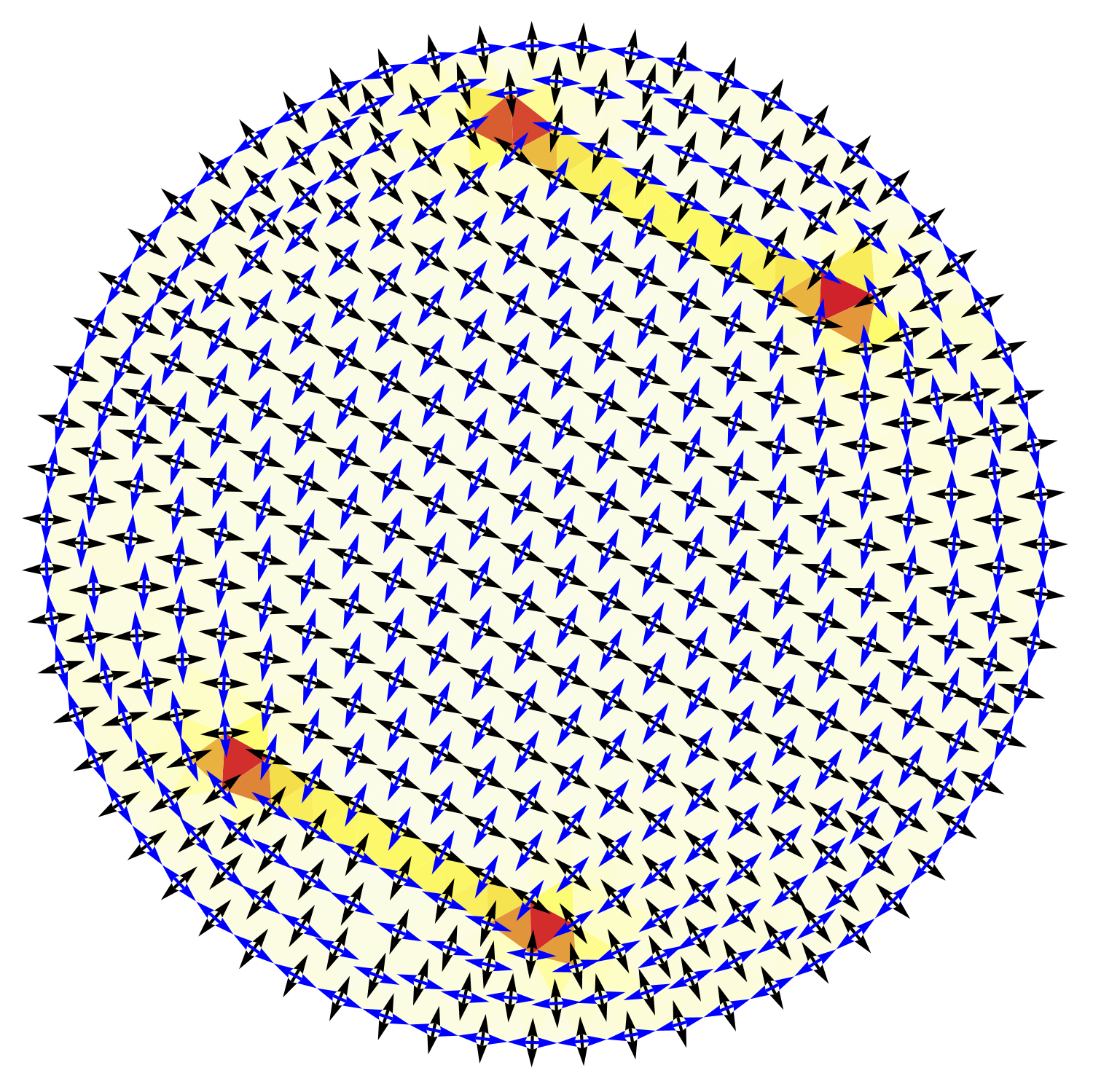}\\
\smallskip
e~\includegraphics[width=.97\mywidth]{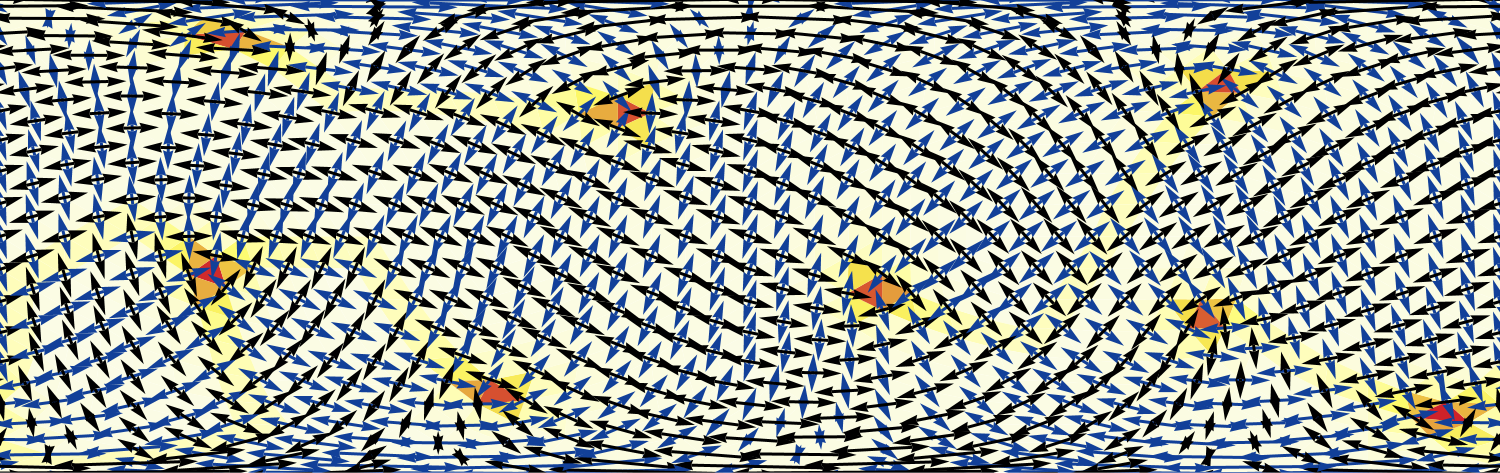}\\
\smallskip
f~\includegraphics[width=.97\mywidth]{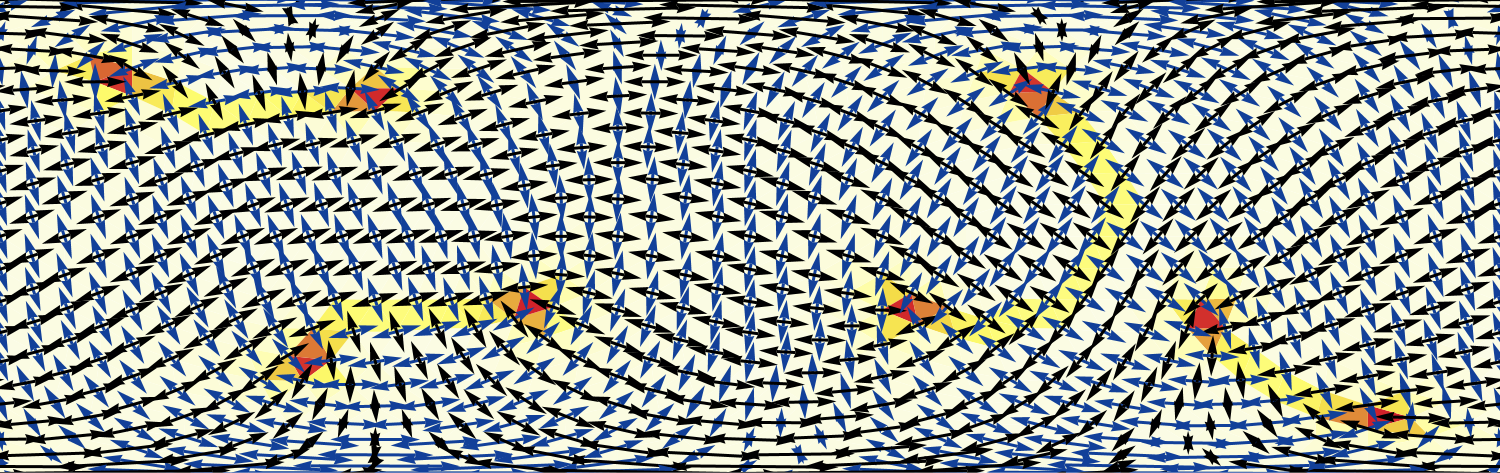}\\
\smallskip
g~\includegraphics[width=.97\mywidth]{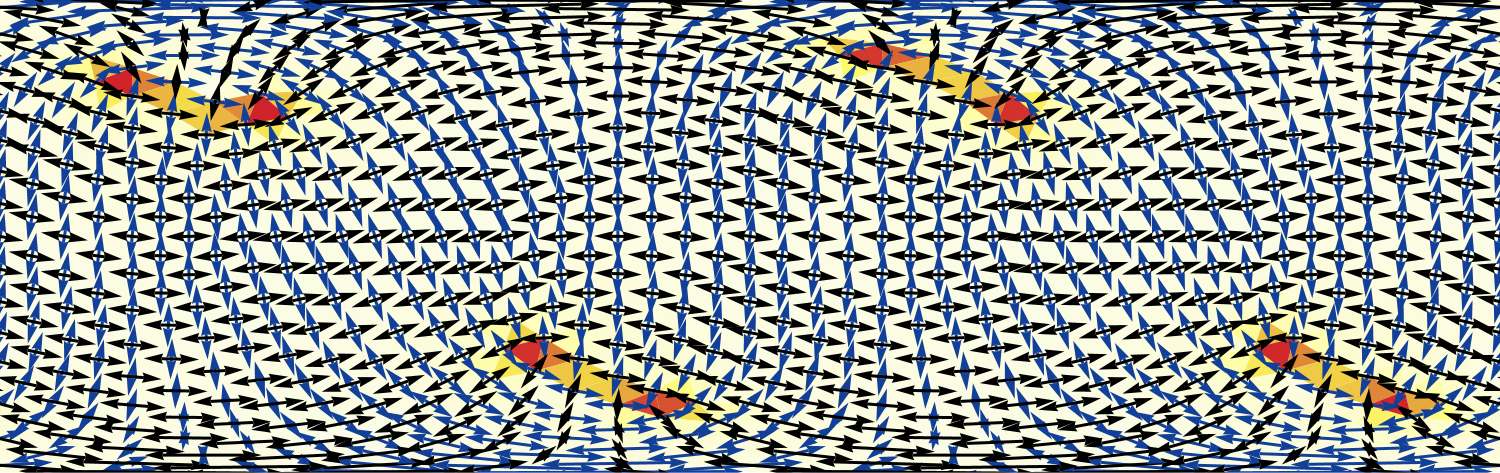}\\
\smallskip
h~\includegraphics[width=.97\mywidth]{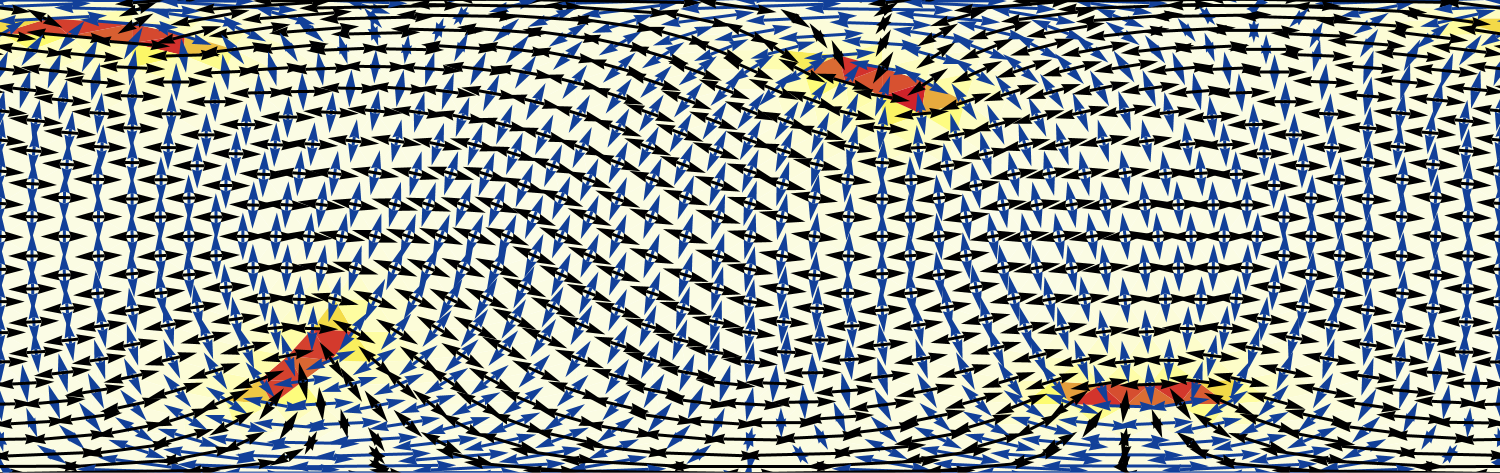}
\caption{Phase with combined nematic and tetratic order.  (a--d) Simulation on disk with $J_4=2$ and $J_2=0.2$, $0.4$, $0.8$, and $1$, respectively.  (e-h) Mercator projection of sphere with $C=5$, $J_4=0.25$, and $J_2=0.05$, $0.1$, $0.2$, and $0.5$.}
\label{fig:nematictetratic}
\end{figure}

Figure~\ref{fig:nematictetratic} presents simulations of a phase with nematic and tetratic order.  Parts (a--d) show the phase on a disk with radial anchoring, and parts (e--h) show the phase on a sphere (in the Mercator projection).  At each mesh site $i$, the orientational order is represented by a $+$ symbol, decorated in black and blue.  The colorless $+$ symbol shows the tetratic order, and the black/blue decoration shows the nematic order.  On the disk, each $+$ symbol is constructed by drawing two black arrows at the angles $\theta_i$ and $\theta_i+\pi$, and two blue arrows at the angles $\theta_i\pm\pi/2$.  Likewise, on the sphere, the black arrows are drawn along the vector $\hat{\bm{s}}_i$, rotated by $0$ and $\pi$ about the local normal.  The blue arrows are drawn along $\hat{\bm{s}}_i$, rotated by $\pm\pi/2$ about the local normal.  The tetratic coupling favors alignment of neighboring $+$ symbols, regardless of color, as if they were four-fold symmetric objects.  The nematic coupling favors alignment of neighboring black with black, and neighboring blue with blue.

As in the previous case of polar and nematic order, the simulations of nematic and tetratic order show point defects with high energy density (red color), as well as domain walls with intermediate energy density (yellow color).  The point defects are topological defects in the tetratic order, with topological charge of $+1/4$ each.  The domain walls are lines where the nematic order reverses, so that the black and blue colors are exchanged, while the tetratic order remains the same.

On the disk, in Figs.~\ref{fig:nematictetratic}(a--d), the tetratic interaction is fixed at $J_4=2$, and the nematic interaction increases from $J_2=0.2$, $0.4$, $0.8$, to $1$.  On the sphere, in Figs.~\ref{fig:nematictetratic}(e--h), the tetratic interaction is $J_4=0.25$, and the nematic interaction increases from $J_2=0.05$, $0.1$, $0.2$, to $0.5$.  For weak nematic interactions, the domain walls are only faint yellow, meaning that they have low energy, and hence they are quite long.  For the smallest ratio $J_2/J_4$, in Figs.~\ref{fig:nematictetratic}(e), we can see that the domain walls form a network connecting the point defects around the sphere.  On both the disk and the sphere, the domain walls occasionally have kinks, which are presumably artifacts of the underlying mesh.

As the nematic interaction increases, the domain walls become brighter yellow, meaning higher energy, and hence they grow shorter.  They bind the point defects into pairs, which can be regarded as string defects.  Each string defect has a total topological charge of $+1/2$.  The disk has two string defects, for a total topological charge of $+1$, as required by the boundary conditions.  The sphere has four string defects, with a total topological charge of $+2$, as required by topology.  For the highest ratio $J_2/J_4$, in Figs.~\ref{fig:nematictetratic}(h), each string shrinks into a single nematic defect with an extended core.

Our results can be compared with recent simulations of colloidal crystals on spherical interfaces~\cite{Jones2025}.  That study uses a very different computational approach from our current paper; it is based on the packing of hard particles with different types of polyhedral shapes.  In their results, the structure and defects depend on the particle shape.  Cubical particles form a square lattice, which has tetratic orientational order as well as crystalline positional order.  Around the sphere, this lattice has eight defects of topological charge $+1/4$ each, similar to our simulation of tetratic order in Fig.~\ref{fig:spheresingleorder}(e).  By contrast, tetrahedral particles form a more complex woven motif, with edges of the tetrahedra oriented inward and outward on the sphere.  In that case, the overall structure on the sphere has many extended defects, which are mostly connected with each other to form a network.  It appears somewhat similar to our simulation of strong tetratic order with weak nematic order in Fig.~\ref{fig:nematictetratic}(e).  We speculate that their results might also be understood in terms of strong tetratic order with weak nematic order.  The woven motif in their simulations certainly has four-fold symmetry, like a tetratic phase.  This four-fold symmetry may be broken down to two-fold, like a nematic phase, by forcing the woven motif into a curved geometry.  The combination of these two types of order may lead to a structure of point defects and domain walls, as we have found.

\subsection{Polar and hexatic}

For a further example of the same type of phenomenon, consider the combination of polar and hexatic order.  One physical realization of this combination is in thin films of smectic liquid crystals~\cite{Nelson1980,Dierker1986,Selinger1988,Selinger1989}. In that case, the hexatic order corresponds to the orientations of bonds between neighboring molecules, which have six-fold symmetry.  The polar order corresponds to the direction of molecular tilt, which has only one-fold symmetry.  These two types of order are coupled, because the tilt generally has a preferred orientation with respect to the local bonds.

\begin{figure}
\centering
a\includegraphics[width=.48\mywidth]{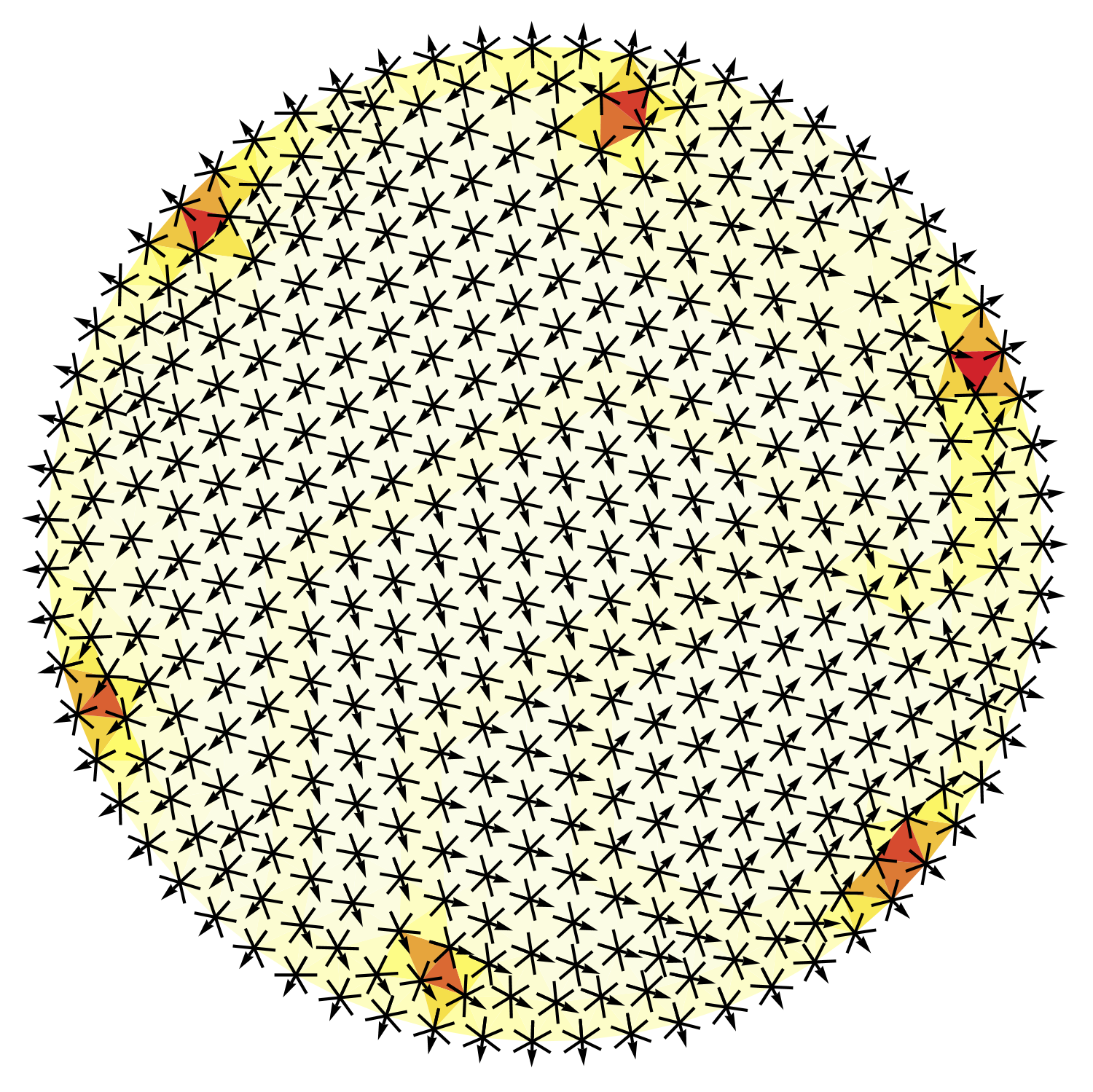}%
b\includegraphics[width=.48\mywidth]{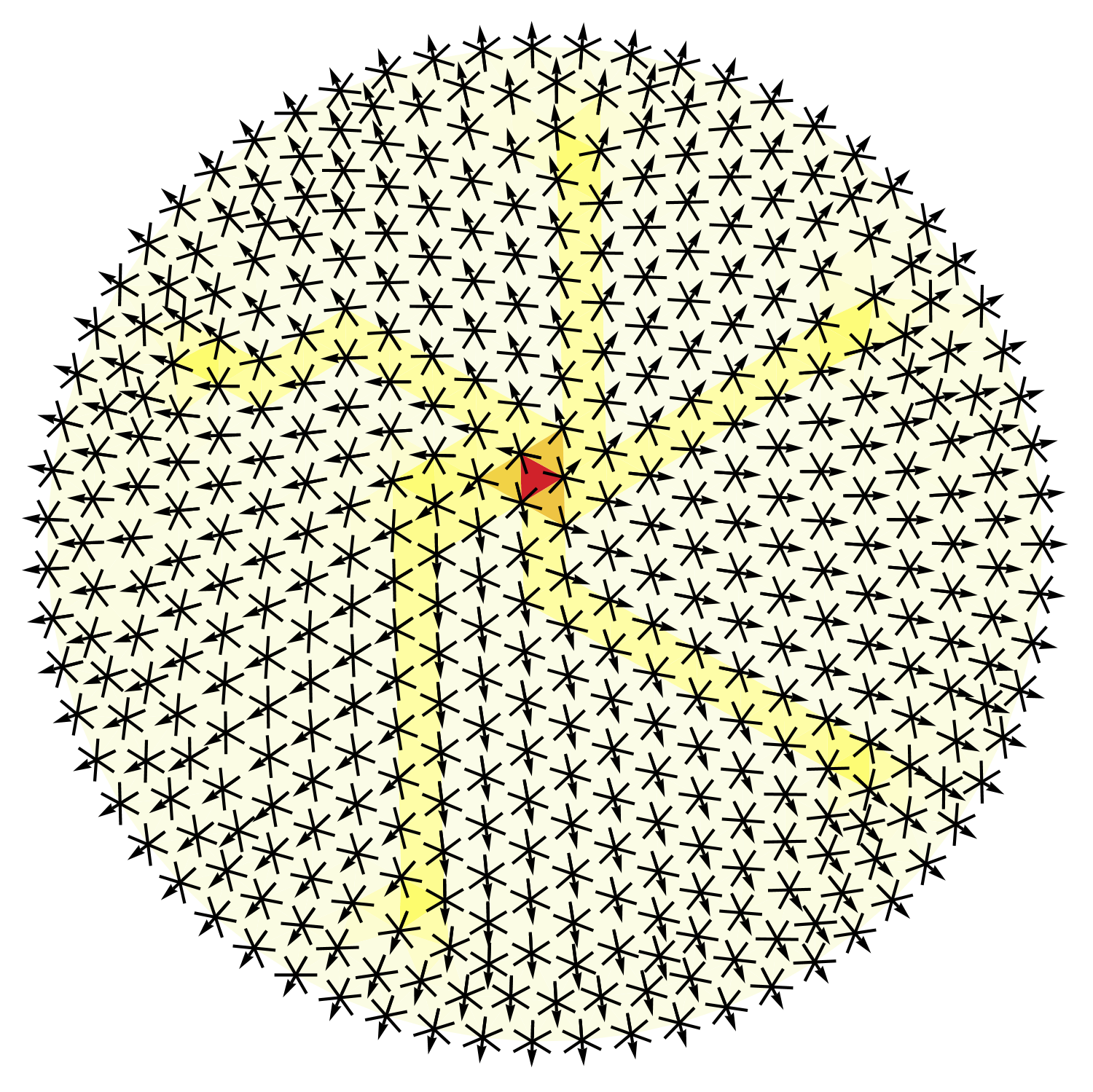}\\
\smallskip
c~\includegraphics[width=.97\mywidth]{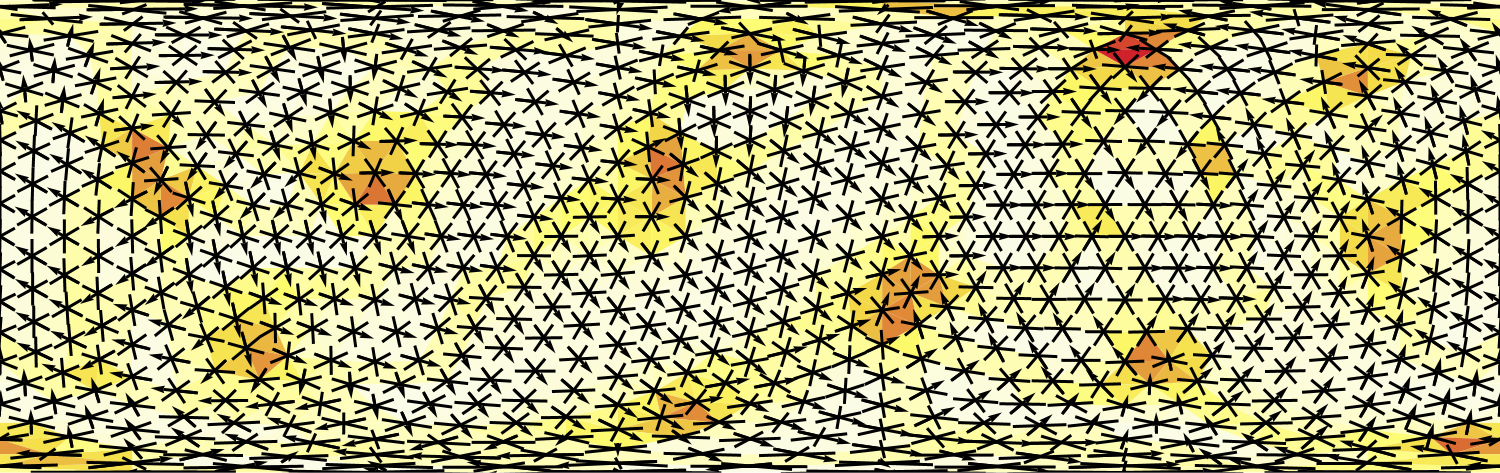}\\
\smallskip
d~\includegraphics[width=.97\mywidth]{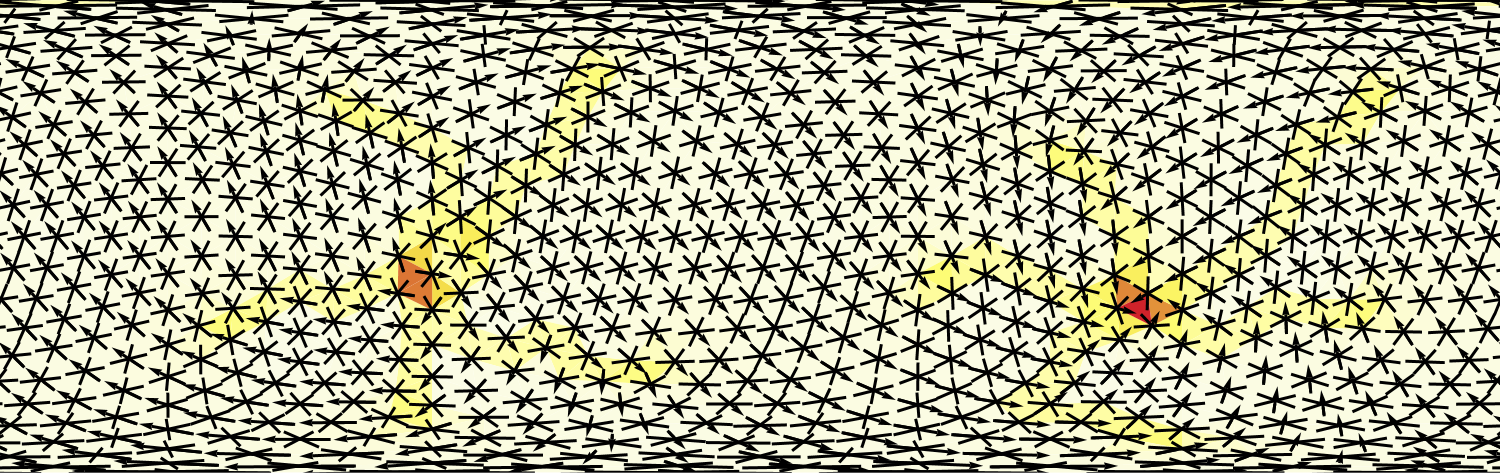}
\caption{Phase with combined polar and hexatic order.  Simulations on disk:  (a)~$J_6=2$, $J_1=0.5$. (b)~$J_6=0.5$, $J_1=2$.  Simulations on sphere: (c)~$C=10$, $J_6=1$, $J_1=0.8$.  (d)~$C=5$, $J_6=0.25$, $J_1=1$.}
\label{fig:polarhexatic}
\end{figure}

Figure~\ref{fig:polarhexatic} shows simulations of the phase with polar and hexatic order, first on a disk with radial anchoring (parts (a--b)) and then on a sphere in the Mercator projection (parts (c--d)).  At each mesh site $i$, the orientational order is represented by a symbol with six arms, decorated by a single arrow.  The six arms show the hexatic order, and the arrow shows the polar order.  For the disk, the arms are drawn at the angles $\theta_i$, $\theta_i+\pi/3$, \dots, $\theta_i+5\pi/3$, and the arrow at $\theta_i$ only.  For the sphere, the arms are drawn along the vector $\hat{\bm{s}}_i$, rotated by $0$, $\pi/3$, \dots, $5\pi/3$ about the local normal, and the arrow along $\hat{\bm{s}}_i$ only.  The hexatic coupling favors alignment of neighboring symbols, regardless of the arrows, as if they were six-fold symmetric objects.  The polar coupling favors alignment of neighboring arrows.

When the hexatic coupling is strong and the polar coupling is weak, as in Figs.~\ref{fig:polarhexatic}(a) and \ref{fig:polarhexatic}(c), the simulations show point defects in the hexatic order, each with topological charge of $+1/6$, which are visible as red points of high energy density.  The disk has six of these defects, for a total topological charge of $+1$, and the sphere has twelve, for a total of $+2$.  The point defects are connected by domain walls, which are visible as yellow lines.  Across each domain wall, the polar order rotates through an angle of $\pi/3$, so that it goes from one of the six directions of hexatic order to another.  The domain walls are fairly low in energy, so they can be quite long, and they connect all of the point defects around the disk or sphere in a large network.

When the polar coupling becomes stronger and the hexatic coupling becomes weaker, as in Figs.~\ref{fig:polarhexatic}(b) and \ref{fig:polarhexatic}(d), the domain walls become much higher in energy.  As a result, the domain walls must become shorter, and hence the system forms ``star defects.''  Each star defect has a point defect at the center, connected to five domain walls as arms, with five more point defects at the ends of the arms.  The point defect at the center is a $+1/6$ defect for the hexatic order, and also a $+1$ defect for the polar order.  For that reason, it has a high energy density, indicated by the red color.  The point defects at the ends of the arms are only $+1/6$ defects for the hexatic orders.  They have lower energy density, and hence only the yellow color.  Each entire star defect has a topological charge of $+1$.  There is one star defect on the disk, and two on the sphere.

The star defects in these simulations are very similar to star defects in tilted hexatic liquid crystals, which were observed experimentally and explained theoretically by Dierker, Pindak, and Meyer~\cite{Dierker1986}.  The main difference is that the experimental structures are more symmetrical, with straighter arms, while our simulated star defects are more disordered, with kinks in the arms.  As in the previous case of nematic and tetratic order, the kinks are presumably artifacts of the underlying mesh.

By comparing the results of this section with the combinations of orientational order in the previous sections, we can see that star defects and string defects are both examples of the same general phenomenon.  We will use the term ``hybrid defects'' to describe the class of topological structures that combine point defects and domain walls in different types of orientational order.

\subsection{Tetratic and hexatic}\label{sec:tetratichexatic}

In all of the examples so far, we have considered a combination of $m$-atic and $n$-atic order where $n$ is a multiple of $m$.  In those cases, the $m$-atic order is a more specific, lower-symmetry version of the $n$-atic order.  In other words, the $m$-atic order selects $m$ special directions among the $n$ special directions associated with $n$-atic order.  It reduces the symmetry by a factor of $n/m$, which is an integer.

As a final example, suppose that a system has a combination of tetratic ($m=4$) and hexatic ($n=6$) order.  In this combination, $n$ is not a multiple of $m$.  The potential energy $H_{4,6}$ of Eq.~(\ref{Hmnflat}) and Fig.~\ref{fig:potentialplots}(d) might have either four or six minima, depending on the ratio $J_4/J_6$.  Two of the minima are located at $\theta_i-\theta_j=0$ and $\pi$.  The locations of the the other minima shift, depending on $J_4/J_6$.

In this case, what should we expect for the hybrid defects?  To answer that question, we note that the greatest common divisor of $m=4$ and $n=6$ is 2, corresponding to nematic order.  For that reason, nematic order is a more specific version of tetratic order, \emph{and} nematic order is a more specific version of hexatic order.  Hence, we speculate that the phase with combined tetratic and hexatic order should form hybrid defects with topological charge of $+1/2$.  Each of these hybrid defects could combine two tetratic defects of charge $+1/4$ each, along with three hexatic defects of charge $+1/6$ each, connected by some combination of domain walls.

\begin{figure}
\centering
a\includegraphics[width=.48\mywidth]{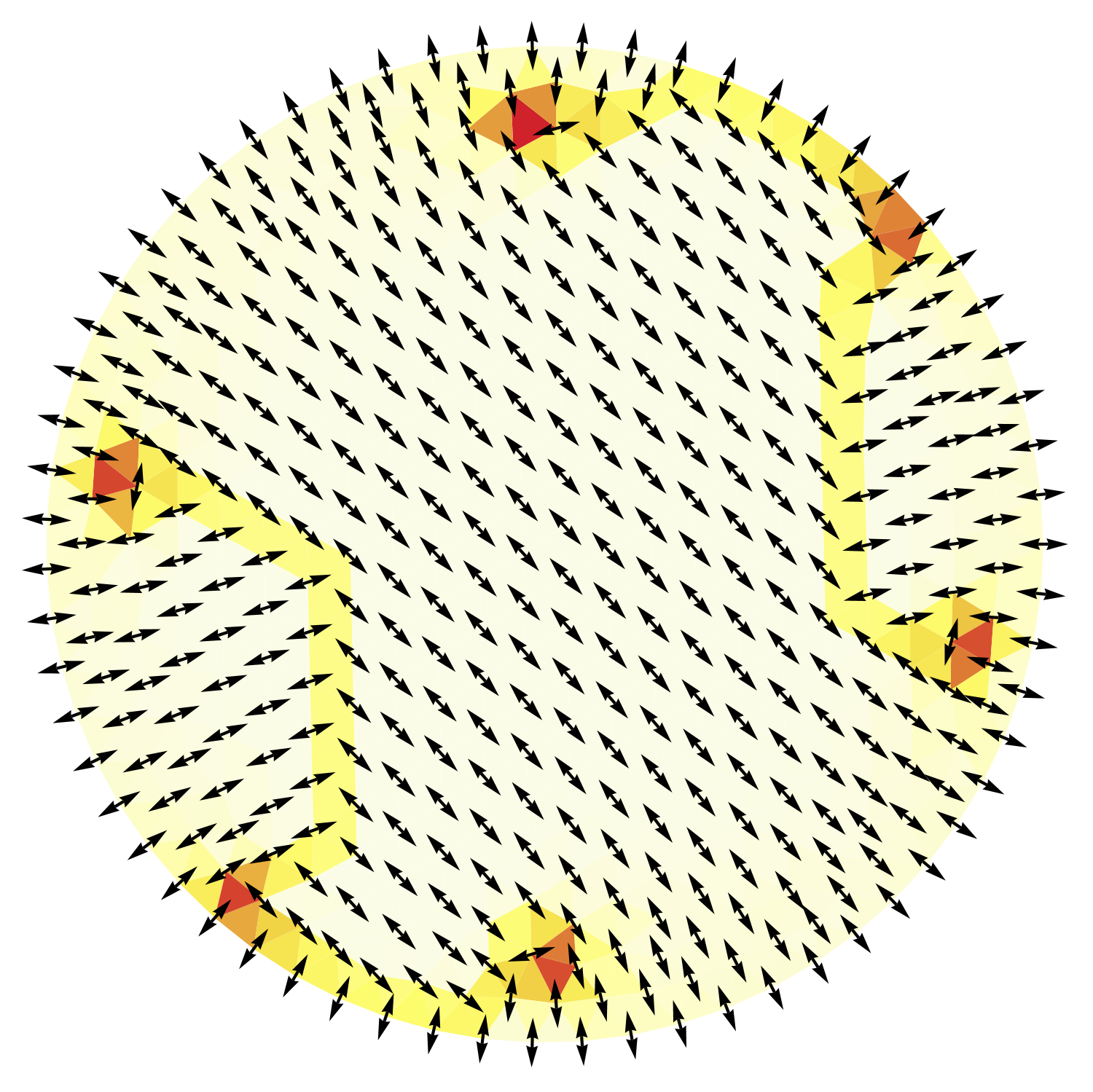}%
b\includegraphics[width=.48\mywidth]{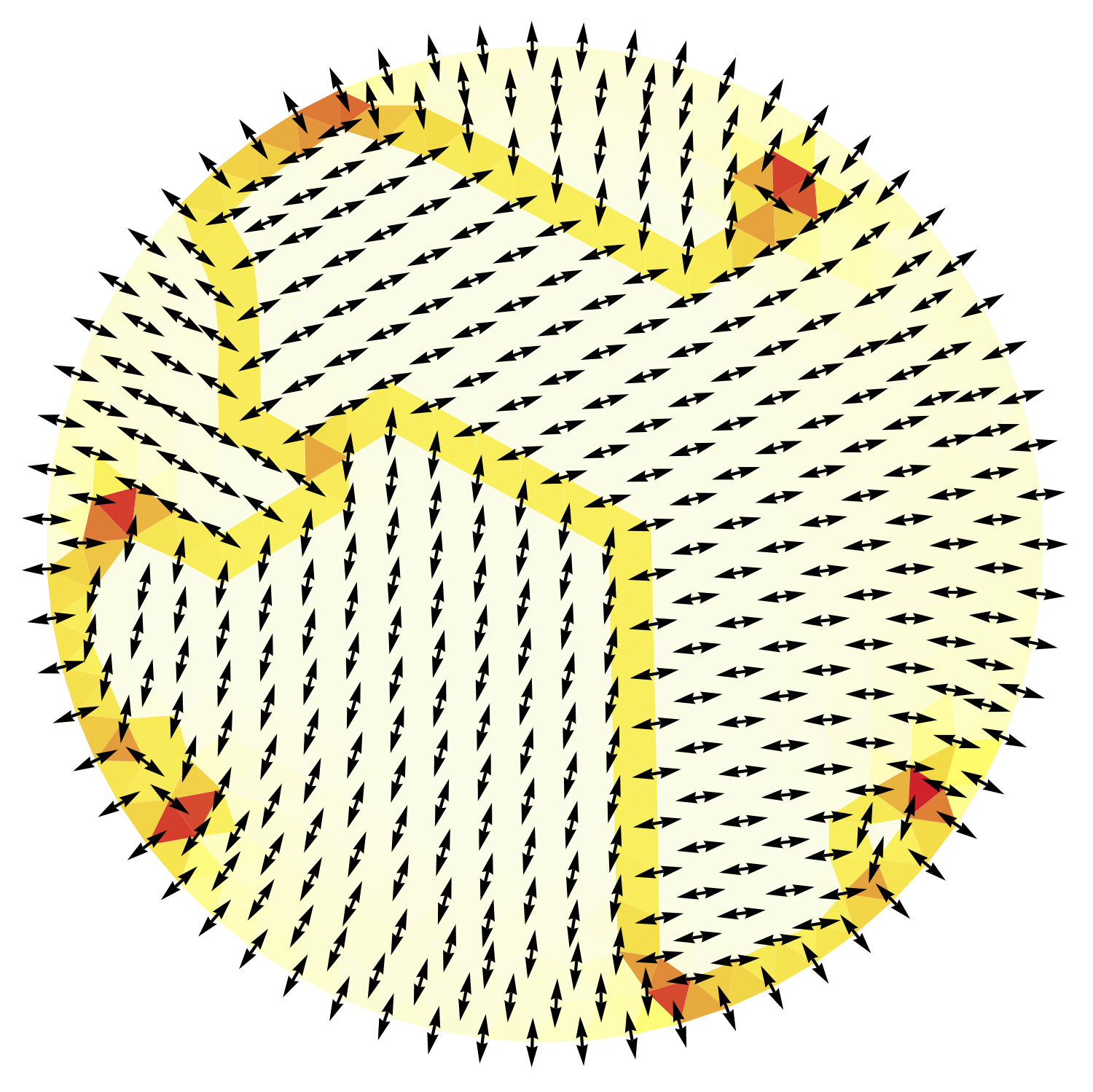}\\
\smallskip
c\includegraphics[width=.48\mywidth]{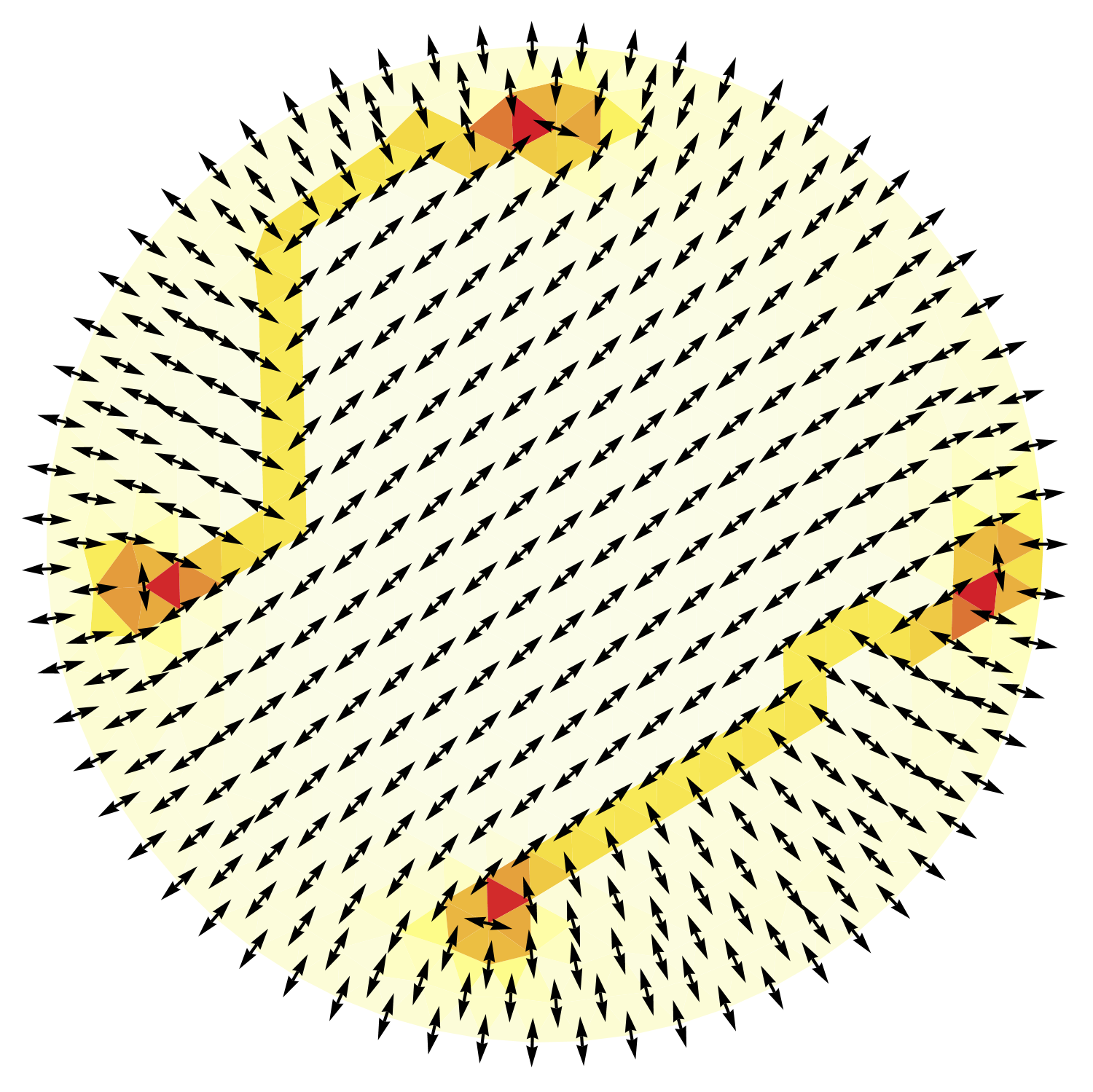}\\
\smallskip
d~\includegraphics[width=.97\mywidth]{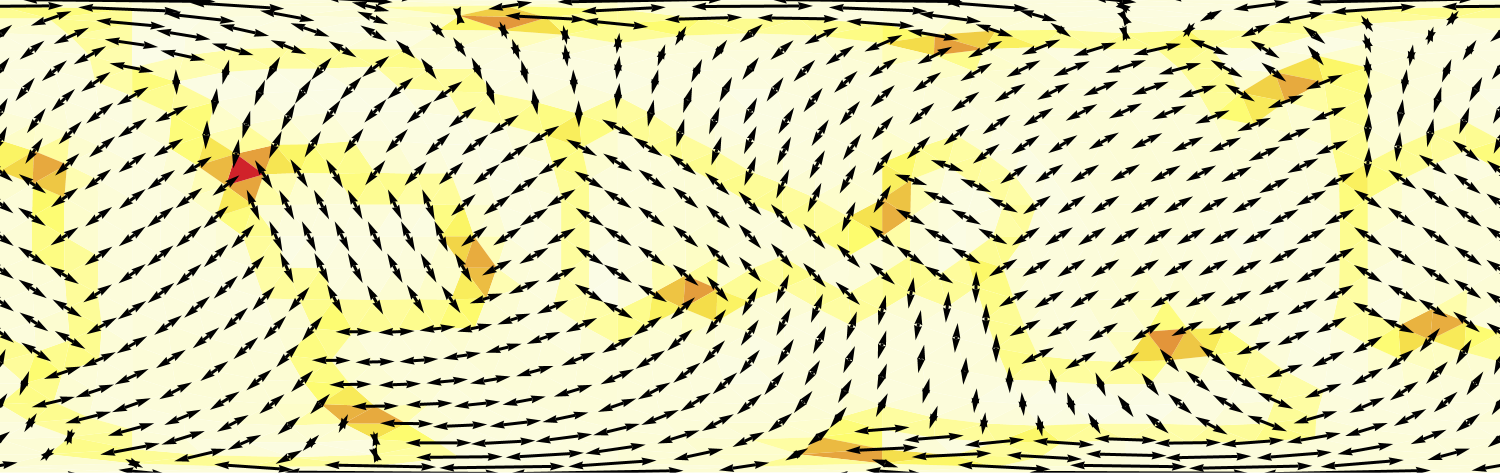}\\
\smallskip
e~\includegraphics[width=.97\mywidth]{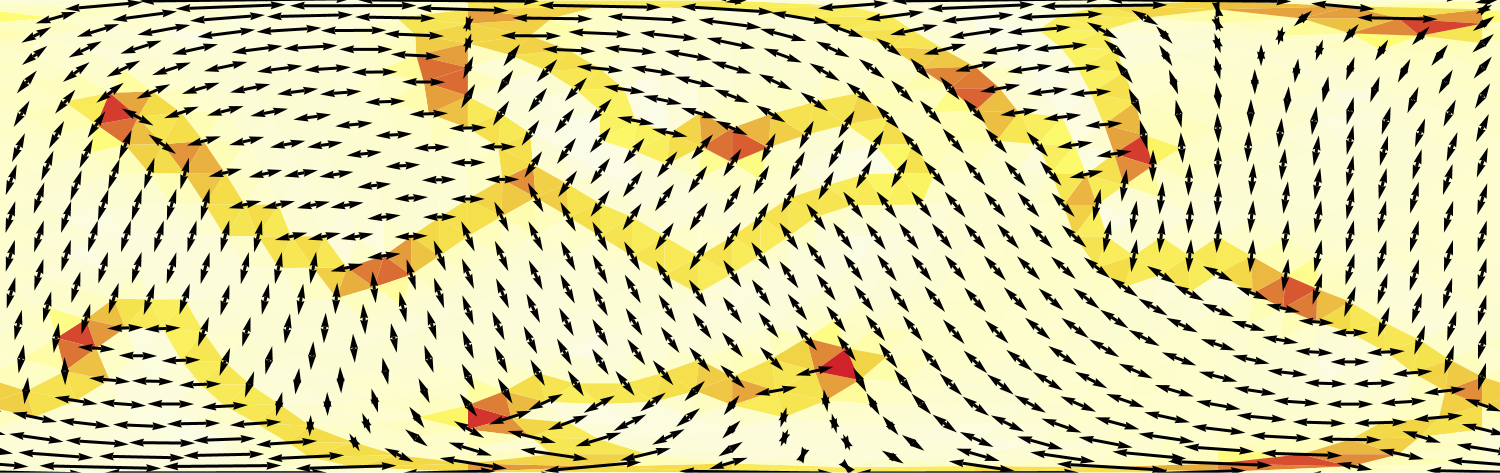}\\
\smallskip
f~\includegraphics[width=.97\mywidth]{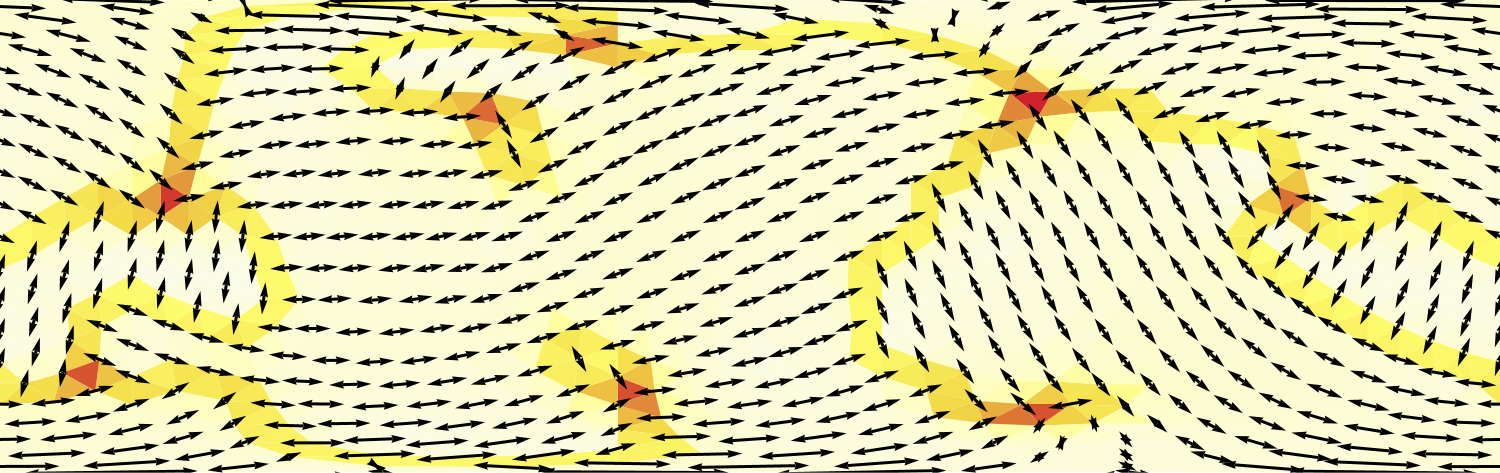}
\bigskip
\caption{Phase with combined tetratic and hexatic order.  Simulations on disk:  (a)~$J_6=2$, $J_4=1$. (b)~$J_6=2$, $J_4=1.5$. (c)~$J_6=2$, $J_4=4$.  Simulations on sphere: (d)~$C=50$, $J_6=1$, $J_4=0.5$.  (e)~$C=50$, $J_6=1$, $J_4=0.75$.  (f)~$C=50$, $J_6=1$, $J_4=2.5$.}
\label{fig:tetratichexatic}
\end{figure}

To test this speculation, we perform simulations of the model with combined tetratic and hexatic order.  Figure~\ref{fig:tetratichexatic}(a--c) shows the results on a disk with radial anchoring, and Fig.~\ref{fig:tetratichexatic}(d--f) on a sphere in the Mercator projection.  At each mesh site $i$, the orientational order is represented by two arrows, which show the angles $\theta_i$ and $\theta_i+\pi$ on the disk, or equivalently the vector $\hat{\bm{s}}_i$ rotated by 0 and $\pi$ about the local normal on the sphere.  We do not draw additional decorations corresponding to tetratic or hexatic order, because they would be too cluttered on the figure.

When the hexatic coupling $J_6$ is large and the tetratic coupling $J_4$ is small, as in Figs.~\ref{fig:tetratichexatic}(a) and~\ref{fig:tetratichexatic}(d), the phase is mainly hexatic, and tetratic order can be regarded as a small perturbation.  Here, the red points of high energy density are hexatic defects of topological charge $+1/6$.  There are six of these points on the disk, and twelve on the sphere.  These points cannot be independent defects, because they are not compatible with the presence of some tetratic order.  Hence, they must be joined by domain walls.  Across each domain wall, the orientational order rotates by approximately $\pi/3$, as expected for a phase that is mainly hexatic.  On the disk, in Fig.~\ref{fig:tetratichexatic}(a), the system forms two hybrid defects, each consisting of three $+1/6$ defects connected by domain walls, each with a total charge of $+1/2$.  However, on the sphere, in Fig.~\ref{fig:tetratichexatic}(c), all of the defects are connected together in a network that spans the entire sphere.

When $J_4$ is large and $J_6$ is small, as in Figs.~\ref{fig:tetratichexatic}(c) and~\ref{fig:tetratichexatic}(f), the phase is mainly tetratic, and the hexatic order is small.  The red points are now tetratic defects of topological charge $+1/4$.  There are four of these points on the disk, and eight on the sphere.  Once again, these points cannot be independent defects, because they are not compatible with the existence of hexatic order, and hence they must be joined by domain walls.  Across each domain wall, the orientational order rotates by approximately $\pi/2$, as expected for a phase that is mainly tetratic.  On the disk, in Fig.~\ref{fig:tetratichexatic}(b), the structure is organized in two hybrid defects, each involving two $+1/4$ defects connected by domain walls, each with a total charge of $+1/2$.  By contrast, on the sphere, in Fig.~\ref{fig:tetratichexatic}(d), all of the defects are again connected together in a network over the whole sphere.

When $J_4$ and $J_6$ are of similar magnitude, as in Figs.~\ref{fig:tetratichexatic}(b) and~\ref{fig:tetratichexatic}(e), the results are more difficult to interpret.  In these cases, the system has a large number of point defects, beyond the minimum number required by topology.  These defects are connected by domain walls, which are longer than the minimum required for the connection.  For both the disk and the sphere, the defects and domain walls form a network over the entire system.  They do not break into hybrid defects with topological charge of $+1/2$.

Overall, the simulation results for tetratic and hexatic order show the same general phenomenon as the results for the other combined order parameters.  As in the previous cases, the system forms point defects linked by domain walls, consistent with the topological requirements for each of the order parameters.  The main difference in this new combination is that the topological structures often cover the entire geometry, instead of breaking into hybrid defects with the expected charge.  It is possible that separate hybrid defects might still be the ground state, but that the simulation algorithm cannot consistently reach this ground state.  We speculate that many arrangements of tetratic and hexatic order have similar energies, and the energy differences are too small to drive relaxation into a simple ground state.

\section{Discussion}

In this paper, we have simulated topological structures that occur when a system has two coupled orientational order parameters, $m$-atic and $n$-atic.  The simulations show that this combination generally results in the formation of hybrid defects, such as the strings and stars found in earlier research.

In the simplest case where $n$ is a multiple of $m$, the hybrid defects involve $n/m$ point defects, each with topological charge $+1/n$.  These point defects are connected by domain walls, across which the orientational order rotates through an angle of $2\pi/n$.  Each of the hybrid defects then has a total topological charge of $+1/m$.

If $n$ is not a multiple of $m$, hybrid defects can form with a topological charge of $+1/l$, where $l$ is the greatest common divisor of $m$ and $n$.  However, in this case, the energy landscape is more complex, and our simulations often show networks of defects extending over the entire geometry.

As we have discussed, combinations of orientational order parameters commonly occur in ferroelectric nematic liquid crystals, tilted hexatic liquid crystals, colloidal crystals on a spherical surface, and one can expect to find them in biological and active materials.  Thus, our results should help to understand topological structures found in this wide range of systems.

\backmatter

\bmhead{Acknowledgements}

We would like to thank H. Aharoni, S. C. Glotzer, G. N. Jones, P. W. A. Sch\"onh\"ofer, and Q.-H. Wei for helpful discussions.


\bibliography{version4}

\end{document}